\documentclass[lettersize,journal]{IEEEtran}
\usepackage[utf8]{inputenc}
\usepackage{amsmath,amsfonts}
\usepackage{algorithmic}
\usepackage{algorithm}
\usepackage{array}
\usepackage[caption=false,font=normalsize,labelfont=sf,textfont=sf]{subfig}
\usepackage{textcomp}
\usepackage{stfloats}
\usepackage{url}
\usepackage{verbatim}
\usepackage{graphicx}
\usepackage{cite}
\usepackage{adjustbox}
\usepackage{colortbl}
\usepackage[table]{xcolor}
\usepackage{multirow}
\usepackage{booktabs}
\usepackage{gensymb}
\usepackage{breqn}
\usepackage{enumitem}
\usepackage{mwe}
\usepackage{caption}
\usepackage{etoolbox, lipsum, babel}

\title{TIDI-GS: Floater Suppression in 3D Gaussian Splatting for Enhanced Indoor Scene Fidelity}

\author{Sooyeun Yang$^{1,2}$, Cheyul Im$^{1,2}$, Jee Won Lee$^{1,2}$, and Jongseong Brad Choi$^{1,2*}$\\
        \footnotesize{$^1$ Department of Mechanical Engineering, State University of New York, Korea, Incheon, South Korea\\
        $^2$ Department of Mechanical Engineering, State University of New York, Stony Brook, NY, United States\\
        \vspace{0.1 cm}
        $^*$ Corresponding Author} 

\thanks{This work was supported by the National Research Foundation of Korea (NRF) grant funded by the Korea government (MSIT) (Grant No. RS-2022-NR067080 and RS-2025-05515607).}}

\IEEEaftertitletext{\vspace{-1.5\baselineskip}
\begin{center}
  \includegraphics[width=\textwidth]{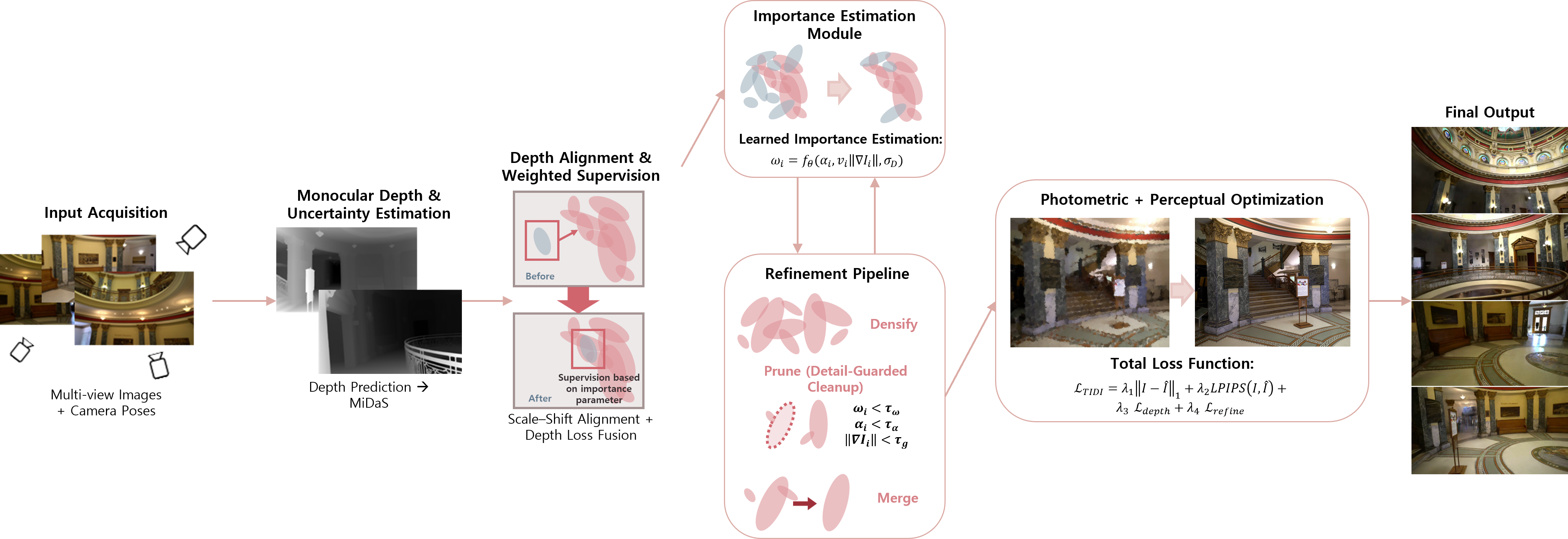}
  \captionof{figure}{Overview of the TIDI-GS framework. Our method introduces two key components to the standard 3DGS pipeline: (left) a soft geometric prior from monocular depth and (center) a periodic, detail-guarded cleanup stage. These modules work in conjunction with the standard densification and optimization steps to produce a refined Gaussian model (right) with suppressed floaters and enhanced geometric fidelity.}
  \label{fig:overview}
\end{center}
\vspace{0.5\baselineskip}} 

\begin{document}

\maketitle


\begin{abstract}
3D Gaussian Splatting (3DGS) is a technique to create high-quality, real-time 3D scenes from images. This method often produces visual artifacts known as \emph{floaters}--nearly transparent, disconnected elements that drift in space away from the actual surface. This geometric inaccuracy undermines the reliability of these models for practical applications, which is critical. To address this issue, we introduce TIDI-GS, a new training framework designed to eliminate these floaters. A key benefit of our approach is that it functions as a lightweight plugin for the standard 3DGS pipeline, requiring no major architectural changes and adding minimal overhead to the training process. The core of our method is a floater pruning algorithm--\emph{TIDI}--that identifies and removes floaters based on several criteria: their consistency across multiple viewpoints, their spatial relationship to other elements, and an importance score learned during training. The framework includes a mechanism to preserve fine details, ensuring that important high-frequency elements are not mistakenly removed. This targeted cleanup is supported by a monocular depth-based loss function that helps improve the overall geometric structure of the scene. Our experiments demonstrate that TIDI-GS improves both the perceptual quality and geometric integrity of reconstructions, transforming them into robust digital assets, suitable for high-fidelity applications.
\end{abstract}


\section{Introduction}
\label{sec:introduction}
\IEEEPARstart{R}{ecent} developments in 3D Gaussian Splatting (3DGS) \cite{3DGS2023} allow interactive, photorealistic 3D scenes in real-time. This is achieved by optimizing millions of small shapes using a spherical rendering process. However, when the viewpoint changes, distracting \emph{floaters} often appear. These artifacts are not just a visual problem; they indicate an unstable geometric structure that undermines the accuracy of measurements, the understanding of object occlusion, and the extraction of surfaces. Even small visual errors can lead to an incorrect view of clearances or defects. This limits the reliability of 3DGS.

This problem is especially challenging in indoor scenes. Capturing indoor spaces involves a mix of bright lights, glossy or glass surfaces, shadows, and large, plain areas such as walls and ceilings. These elements provide weak visual cues from multiple views and can confuse the optimization process. 3DGS may create translucent Gaussians in empty space to explain these view-dependent effects. Traditional cleanup methods that remove elements based on their transparency frequently fail to eliminate these artifacts. They can also accidentally remove genuine thin structures such as wires or edges. Previous research has mainly focused on outdoor or simpler scenes, leaving the challenges of indoor environments unresolved. 

We hypothesize that an effective floater-removal strategy should be: (i) evidence aware: it should track an object's visibility from multiple views over time, rather than just relying on its transparency at one moment, (ii) context aware: it should identify if a point is isolated in space by examining its neighbors, and (iii) detail aware: it should protect fine details using specific visual cues. Our TIDI-GS (\textbf{T}raining for \textbf{I}ndoor scenes with \textbf{D}etail-aware pruning and \textbf{I}mportance-weighting for \textbf{G}aussian \textbf{S}platting) applies these principles. It adds a regular cleanup stage to the standard training process that is driven by evidence and context, similar to Micro-Splatting \cite{MicroSplatting2024}. It also uses depth information to help stabilize the geometry without requiring major architectural changes or causing significant overhead. 

Our evaluation is designed to highlight its practical benefits for interactive applications rather than the algorithm itself. While we report standard image quality metrics such as PSNR (Peak Signal-to-Noise Ratio), SSIM (Structural Similarity Index Measure), and LPIPS (Learned Perceptual Image Patch Similarity), we focus our analysis on performance aspects that directly capture the visual problems caused by floaters. We assess improvements in silhouette integrity, background consistency, and depth stability during camera motion. Our experiments on various indoor scenes show that TIDI-GS reduces visual noise and errors without sacrificing the fidelity of thin structures. Further tests confirm the contribution of each component in our system. We also show that our cleanup process adds very little time to the standard training. These results indicate that TIDI-GS is a practical upgrade for creating inspection-grade indoor models. TIDI-GS precisely targets floaters, removing unsupported Gaussians while protecting real details, and it integrates easily into existing 3DGS pipelines. 

Our main contributions are:
\begin{itemize}[leftmargin=*]
    \item A method that combines evidence from multiple views and spatial isolation, going beyond simple transparency-based cleaning or complex architectural modifications.
    \item A complete training framework that includes a depth regularization loss that accounts for uncertainty to produce geometrically accurate, visually clean results.
    \item A comprehensive evaluation on challenging indoor scenes, using analysis targeted at the geometric stability issues common in inspection scenarios.
\end{itemize}

\section{Related Work}
\label{sec:related work}
\subsection{From Neural Radiance Fields to 3D Gaussian Splatting}

Novel View Synthesis was significantly advanced by Neural Radiance Fields (NeRF) \cite{NeRF2020}, which introduced an implicit volumetric representation of a scene. However, NeRF's requirement for dense sampling during rendering leads to slow performance. Much research explored explicit and hybrid representations, such as multi-resolution grids \cite{InstantNGP2022}, to accelerate rendering. 3D Gaussian Splatting (3DGS) \cite{3DGS2023} has emerged as a leading method for real-time photo-realistic rendering. It represents scenes with millions of anisotropic Gaussians and uses a tile-based rasterizer for efficiency. This approach is based on ideas from classic point-based rendering techniques \cite{SurfaceSplatting2001, EWASplatting2002, QSplat2000}. However, a significant disadvantage of the standard optimization process is the creation of distracting floater artifacts, which degrade the reconstruction's geometric fidelity. While recent work like Micro-Splatting \cite{MicroSplatting2024} has improved the stability of the Gaussians, it does not directly address the problem of \emph{floater} artifacts.

\subsection{Controlling Geometry and Artifacts in 3DGS}

A key challenge in 3DGS is managing the density of Gaussians to accurately represent surfaces without introducing artifacts. The standard 3DGS training loop uses simple and view-independent rules for pruning, such as removing Gaussians with low opacity \cite{3DGS2023}. This approach is often insufficient, as it can preserve unsupported floaters in empty space while incorrectly removing legitimate thin structures. Methods like Pixel-GS \cite{PixelGS2024} use pixel-aware gradients to guide the creation of Gaussians more carefully, which can reduce artifacts but at the cost of increased model size and training time. An effective pruning strategy requires a more subtle approach: one that accumulates evidence over multiple views \cite{3DGS2023}, understands the spatial context to identify isolated points \cite{PixelGS2024}, and includes explicit safeguards to preserve high-frequency details \cite{MicroSplatting2024}.

To further improve geometric accuracy, especially in indoor scenes with large or textureless regions, external geometric priors can be used. This concept was demonstrated in the context of NeRF, where supervising the training with even noisy depth maps was shown to improve results \cite{DepthSupervisedNeRF2022}. When ground truth depth is not available, monocular depth estimators such as MiDaS \cite{MiDaS2022, DPT2021} and Depth Anything \cite{DepthAnything2024, DepthAnythingV2_2024} can provide valuable, although scale and shift ambiguous, geometric cues. In our work, we integrate these priors as a soft constraint, using an uncertainty-aware loss function that adapts its influence during training \cite{UncertaintyWeighting2018}. This helps to prevent the initial formation of floaters without relying on hard pruning from potentially biased depth estimates.

\subsection{Evaluating Scene Fidelity}

While standard metrics like PSNR, SSIM, and LPIPS \cite{LPIPS2018} are widely used to measure image quality, they do not fully capture the interactive failure modes common to 3DGS. Artifacts like noisy floaters, incorrect silhouette boundaries, and depth inconsistencies under camera motion are often missed by these static metrics \cite{MipNeRF2021}. To provide a more complete picture of reconstruction quality for practical applications, recent work has complemented these scores with analysis that directly measures geometric stability. Our evaluation includes indicators of silhouette integrity, depth stability under small viewpoint changes, and background consistency to better assess the cleanliness and reliability of the final model.


\section{Preliminaries}
\label{sec:preliminaries}

Our work builds upon the 3D Gaussian Splatting (3DGS) framework. This section briefly reviews the core rendering mechanism and introduces the key concepts we use for identifying floaters and regularizing geometry.

\subsection{Differentiable Rasterization in 3DGS}

In 3DGS, a scene is represented by a large set of 3D Gaussians, each defined by properties such as its position, shape (covariance), opacity, and color, which is modeled using Spherical Harmonics (SH) to capture view-dependent effects \cite{3DGS2023}. To create an image, these 3D Gaussians are projected onto the 2D image plane. The final color of each pixel is then calculated by blending the projected Gaussians in order from front to back. This entire process is differentiable, which allows the properties of all Gaussians to be optimized using gradient-based methods. The final color $\hat{\mathbf{I}}(\mathbf{u})$ for a pixel $\mathbf{u}$ is determined by the $\alpha$-blending equation: 

\begin{equation}
    \hat{\mathbf{I}}(\mathbf{u}) = \sum_{k} T_k(\mathbf{u}) \alpha_k \mathbf{c}_k
\end{equation}

$\mathbf{c}_k$ represents the color of the $k$-th Gaussian as determined by its SH coefficients, and $\alpha_k$ is its opacity. The term $T_k(\mathbf{u})$ denotes the transmittance, which calculates how much light from Gaussians behind the $k$-th one can reach the camera. It is computed by multiplying the capacities of all Gaussians in front of it, ensuring that closer objects correctly occlude those farther away.

\subsection{Evidence Accumulation for Floater Identification}

A common artifact in 3DGS is the creation of \emph{floaters}, which are unsupported Gaussians that appear to float in empty space. To distinguish these from legitimate parts of the scene, we track several lightweight statistics for each Gaussian during the training process. These statistics provide evidence to guide our pruning algorithm. 

First, we track multi-view visibility by counting the number of training views in which a Gaussian significantly contributes to the rendered image. A low count suggests the Gaussian is not part of a consistent surface. We also monitor optimization activity using a moving average of a Gaussian's position gradient. A small gradient means the optimizer has less reason to move the Gaussian, suggesting it might no longer be needed. We introduce a learnable scalar weight for each Gaussian as a measure of its learned importance, allowing the model to determine which Gaussians are most critical for minimizing the loss. We employ several detail-preserving guards to prevent the accidental removal of fine details. These guards protect Gaussians that exhibit signs of real structure, such as high-frequency color information, significant local texture, or thin, elongated shapes. We measure spatial isolation by calculating a Gaussian's average distance to its nearest neighbors. Floaters typically have a large isolation distance, whereas Gaussians representing real surfaces are found in dense clusters. These signals enable us to differentiate unsupported artifacts from legitimate scene geometry \cite{3DGS2023, PixelGS2024}.

\subsection{Geometric Priors from Monocular Depth}

In indoor scenes that often contain large, textureless surfaces, relying on color information can be insufficient for reconstructing accurate geometry. We address this by incorporating a geometric prior from a pretrained monocular depth estimation network, MiDaS \cite{MiDaS2022, DPT2021}. We align the rendered depth with the predicted depth for each camera view because these depth predictions are only accurate up to an unknown scale and shift. We then introduce a loss term that is weighted by the uncertainty of the depth prediction, which is a principle derived from multi-task learning \cite{UncertaintyWeighting2018}. 

\begin{equation}
    \mathcal{L}{\text{depth}} = \sum{p} w_{\text{uncert}}(p) \cdot \rho!\big(s,\hat{Z}(p)+t-\widetilde{Z}(p)\big)
\end{equation}

For each pixel $p$, $s$ and $t$ are the scale and shift parameters used to align the rendered depth $\hat{Z}(p)$ with the monocular prediction $\widetilde{Z}(p)$. $w_{\text{uncert}}(p)$ is a crucial uncertainty weight that reduces the loss contribution from pixels where the depth estimator is likely to be inaccurate (e.g., on reflective surfaces or object boundaries). $\rho$ is a Huber loss function, which is less sensitive to large prediction errors. This approach provides soft geometric guidance that discourages the formation of floaters during training and complements our evidence-based cleanup stage, which removes any floaters that still appear. 


\section{Methodology}
\label{sec:methodology}

TIDI-GS enhances the standard 3D Gaussian Splatting (3DGS) framework by introducing two complementary mechanisms. The primary component is a periodic cleanup process that systematically identifies and removes floaters based on accumulated evidence. This is paired with a lightweight monocular depth regularizer that guides the geometry toward more likely surfaces during training, reducing the initial formation of such artifacts. Both components are designed as plugin modules that operate within the standard 3DGS training loop without changing its core architecture, thereby preserving its real-time rendering capabilities and adding minimal computational overhead \cite{3DGS2023, PixelGS2024}.

\subsection{Evidence Accumulation and Candidate Pooling}

\begin{figure}[h!]
    \centering
    \includegraphics[width=\linewidth]{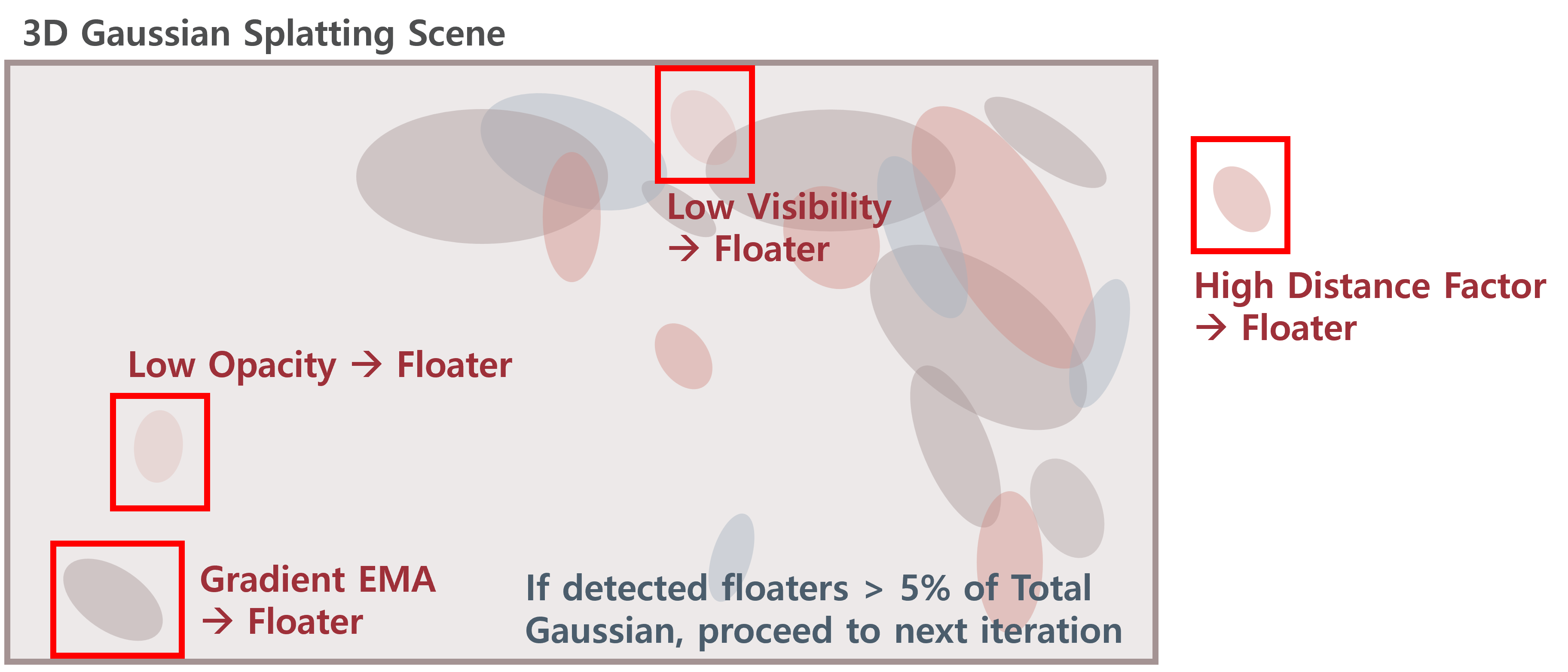}
    \caption{Conceptual illustration of floater cues. Our method identifies floaters based on a combination of weak evidence signals, including low opacity, low multi-view visibility, low optimization activity (Gradient EMA), and high spatial isolation (High Distance Factor). These signals are used to trigger a periodic pruning pass to clean the scene representation. }
    \label{fig:floater_detection}
\end{figure}

Our method maintains an \emph{evidence log} for each Gaussian that accumulates multiple cues over time, enabling more informed and adaptive pruning decisions instead of relying on instant opacity thresholds. In Fig \ref{fig:floater_detection}, a Gaussian is considered a potential floater if it exhibits multiple and persistent signs of being unsupported. We track four primary cues to build this evidence. 

The first cue is \emph{multi-view support} that is measured by a running visibility counter $v_i$. This counter matches the number of training iterations in which a Gaussian $g_i$ makes a non-trivial contribution to any rendered pixel. A low count over an extended period indicates that the primitive is not part of a coherent surface observed from multiple viewpoints, but rather an artifact that may only explain a local residual in a few specific views. We monitor \emph{optimizer usage} through an exponential moving average (EMA) of the position-gradient norm. This metric tracks whether the optimizer is actively relocating the Gaussian to reduce the training loss. The update rule is: 

\begin{equation}
    g_i^{(t)}=\beta g_i^{(t-1)}+(1-\beta)||\nabla_{\mathbf{x}_i}\mathcal{L}||_2
\end{equation}

where $g_i^{(t)}$ is the EMA at training step, $t$ and $\beta$ is a decay rate that smooths out the signal over time, and $||\nabla_{\mathbf{x}_i}\mathcal{L}||_2$ is the magnitude of the gradient of the loss $\mathcal{L}$ with respect to the Gaussian's center $\mathbf{x}_i$. A continuously small EMA value suggests the Gaussians have become a passenger in the optimization and contribute little to further error reduction. Next, a learnable scalar parameter $\omega_i$ provides a measure of \emph{learned importance}, allowing the model itself to assign utility to each primitive. Each scale $\omega_i$ is initialized as a model parameter (e.g., within a value of 1.0) and is optimized jointly with other Gaussian attributes such as position and color through standard back-propagation. This allows the optimizer to dynamically increase the importance of primitives that are decisive for reducing the loss and decrease it for those that are redundant or harmful. We also consider its \emph{opacity} $\alpha_i$ as a measure of radiometric contribution. 

\begin{figure*}[t]
  \centering
  \includegraphics[width=\textwidth]{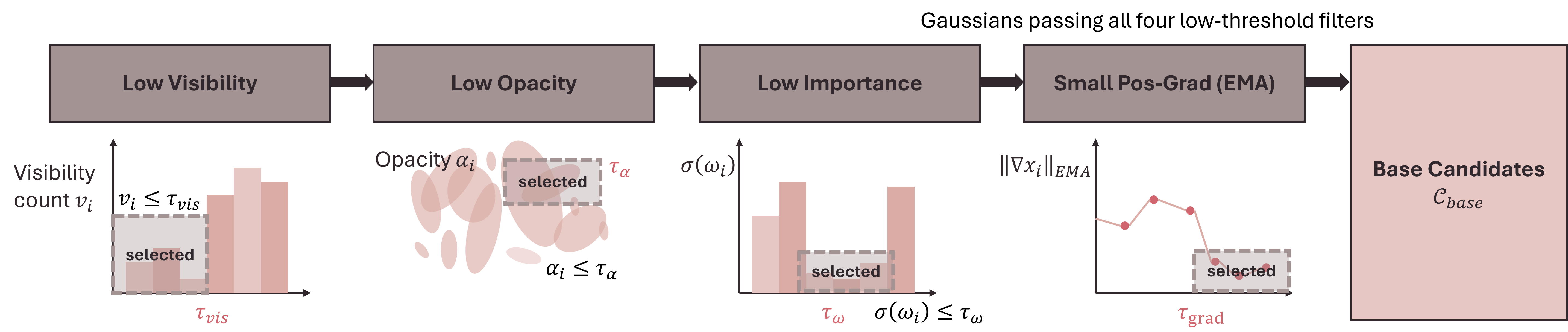}
  \caption{The pipeline for identifying the base candidate set $\mathcal{C}_{\text{base}}$ for pruning. A Gaussian is selected for this set if it falls below predefined thresholds for all four weak evidence signals shown: (i) its visibility count ($v_i$) indicates insufficient multi-view support; (ii) its opacity ($\alpha_i$) shows a negligible radiometric contribution; (iii) its learned importance ($\sigma(\omega_i)$) is low; and (iv) its position-gradient EMA ($||\nabla x_i||_{\text{EMA}}$) is small. This multi-criteria approach creates an initial set of potential floaters that will be further refined by the detail-preserving guards. }
  \label{fig:thresholds}
\end{figure*}

The pipeline in Fig. \ref{fig:thresholds} shows how these cues are combined to form a base candidate set $\mathcal{C}_{\text{base}}$. We flag any Gaussian that falls below predefined, conservative thresholds for all four weak-evidence signals $(\tau_{\text{vis}}, \tau_\alpha, \tau_\omega, \tau_{\text{grad}})$ during training. This initial step functions as a coarse-grained filter and is intentionally excessive to ensure that all potential floaters are captured for more detailed analysis in the subsequent stages. 


\subsection{Detail Preserving Guards}

Aggressive pruning of the base candidate set could unintentionally destroy legitimate fine structures that may share characteristics with floaters like low opacity. Therefore, we apply a series of \emph{detail preserving guards} that function as a test for visually significant information. A candidate Gaussian is exempted from the pruning process if it passes any of these checks. One set of guards is related to appearance, assessing features such as high-frequency content encoded in the spherical harmonics (SH) coefficients. The energy of the non-DC (Direct Current) terms $E_i^{SH}$ is a strong indicator of view-dependent effects like highlights. Such effects are critical for photorealism and must be preserved \cite{RefNeRF2022}. We also measure local color variance to identify Gaussians in textured regions. 

Guards for geometry analyze the Gaussian's shape as derived from its 3D covariance matrix. We compute its thinness and anisotropy from its scales $\left( s_{i, 1}\leq s_{i, 2}\leq s_{i,3}\right)$: 

\begin{equation}
    \mathcal{X}^{\text{thin}}i=s{i, 1}, \quad \mathcal{X}i^{\text{aniso}}=\dfrac{s{i, 3}}{s_{i, 1}}
\end{equation}

A very small scale $\mathcal{X}_i^{\text{thin}}$ signifies a wire-like object, while a high anisotropy ratio $\mathcal{X}_i^{\text{aniso}}$ indicates a flat or planar structure. Both often represent critical geometric details that standard opacity-based pruning might mistakenly remove \cite{3DGS2023}. The thresholds for these guards are set adaptively based on the distribution of features within the population of stable, non-candidate Gaussians, making the method adaptable to different scene scales and content. Only candidates that fail all of these guard tests proceed to the final removal stage. This ensures that our method conservatively prioritizes preserving important details. 

\begin{algorithm}[h]
\caption{Detail-Aware Floater Pruning}
\label{alg:pruning}
\begin{algorithmic}[1]
\STATE \textbf{Input:} All Gaussians $\mathcal{G}$, thresholds $\tau_{\dots}$
\STATE \textbf{Output:} Pruned set of Gaussians $\mathcal{G}'$
\STATE
\STATE // Step 1: Identify Initial Candidates
\STATE $\mathcal{C}_{\text{base}} \leftarrow \{ g_i \in \mathcal{G} \mid v_i \le \tau_{\text{vis}} \land \alpha_i \le \tau_{\alpha} \land \sigma(\omega_i) \le \tau_{\omega} \land \|\nabla_{\mathbf{x}_i}\|_{\text{EMA}} \le \tau_{\text{grad}} \}$
\STATE
\STATE // Step 2: Apply detail-preserving Guards
\STATE $M_{\text{detail}} \leftarrow \{ g_i \in \mathcal{C}_{\text{base}} \mid \|\mathbf{f}_{i, \text{rest}}\|_2 \ge \tau_{H} \lor V_i \ge \tau_V \lor \min(\mathbf{s}_i) \le \tau_s \}$
\STATE $\mathcal{C}_{\text{prune}} \leftarrow \mathcal{C}_{\text{base}} \setminus M_{\text{detail}}$
\STATE
\STATE // Step 3: Isolation-Based Removal with Adaptive Capping
\STATE $\mathcal{G}_{\text{candidates\_to\_remove}} \leftarrow \{ g_i \in \mathcal{C}_{\text{prune}} \mid d_i \ge \tau_{\text{iso}} \cdot \text{extent} \}$
\STATE $\mathcal{G}_{\text{removed}} \leftarrow \text{TopK}(\mathcal{G}_{\text{candidates\_to\_remove}}, \text{cap\_ratio})$
\STATE $\mathcal{G}' \leftarrow \mathcal{G} \setminus \mathcal{G}_{\text{removed}}$
\STATE \textbf{return} $\mathcal{G}'$
\end{algorithmic}
\end{algorithm}


\subsection{Isolation Aware Pruning with Adaptive Caps}

The final set of unguarded candidates is pruned based on its spatial context. This step is founded on the assumption that valid geometric structures form dense or coherent manifolds, whereas floaters are weakly connected to these structures and thus spatially isolated \cite{PixelGS2024} by definition. We quantify this isolation by calculating the average distance $d_i^{(k)}$ of each candidate to its $k$-nearest neighbors (k-NN) in 3D space. A composite pruning score is then computed for each candidate, and it combines its normalized spatial isolation with other weak evidence signals like low opacity and low learned importance. This score provides a ranking of the candidates, and it prioritizes those that are simultaneously isolated, transparent, and considered unimportant by the optimizer. A greedy pass then removes candidates with the highest scores. 

\begin{figure}[h!]
    \centering
    \includegraphics[width=0.7\linewidth]{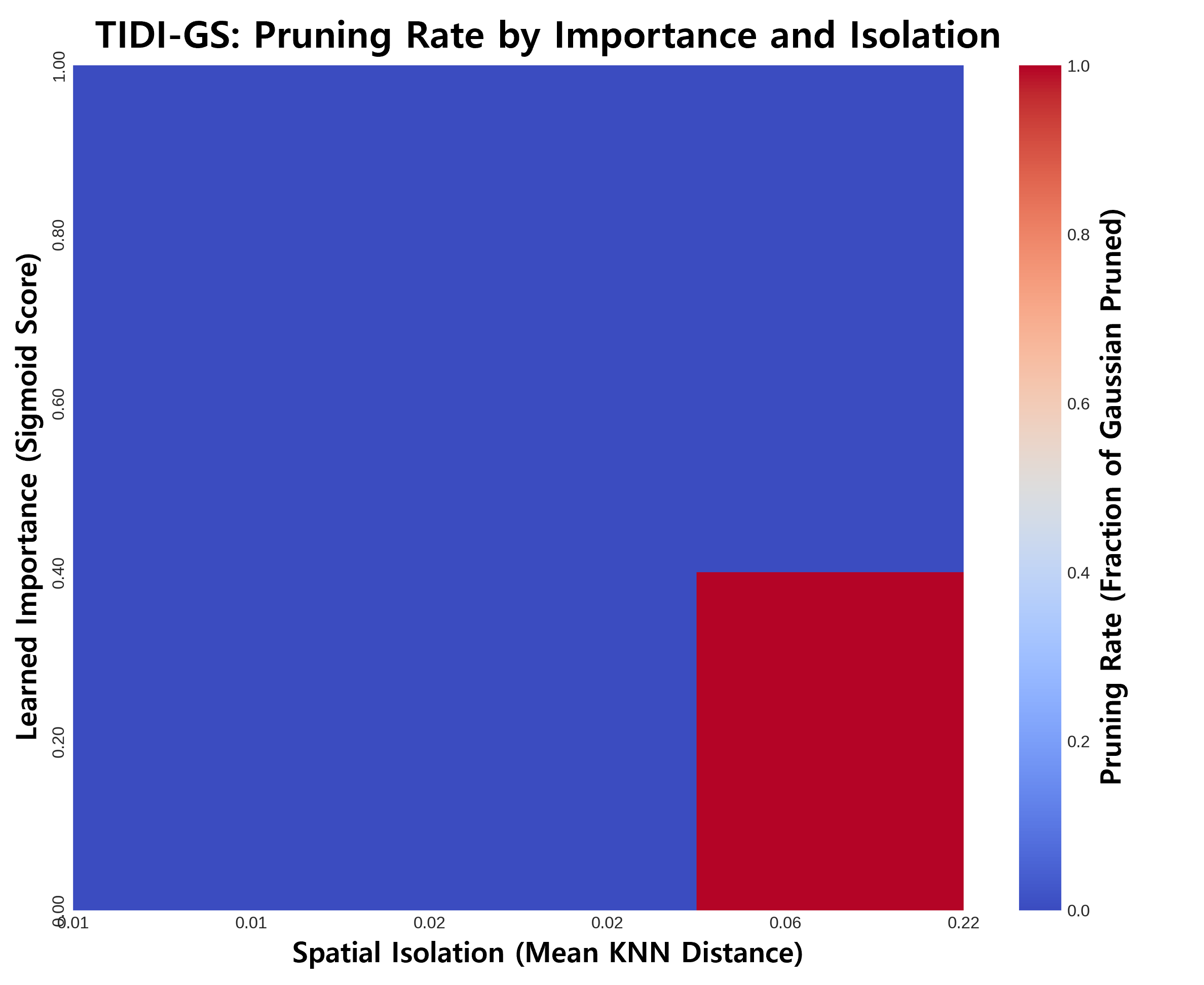}
    \caption{Empirical pruning rate as a function of spatial isolation and learned importance. The high pruning rate (red) is concentrated in the region of high spatial isolation (large $k$-NN distance) and low learned importance, confirming our method's targeted removal of floaters. }
    \label{fig:heatmap}
\end{figure}

Fig. \ref{fig:heatmap} provides empirical validation for this strategy. The heatmap shows the fraction of pruned Gaussians as a function of their spatial isolation (x-axis) and learned importance (y-axis). The red region, which indicates a high pruning rate of nearly 100\%, is concentrated in the area of high spatial isolation and low learned importance. This confirms that our removal policy has learned to precisely target the population of Gaussians that are characteristic of floaters. To maintain training stability and prevent excessive removals in a single step, we impose adaptive local (per-cell) and global caps on the number of Gaussians pruned. This ensures that the cleanup is balanced and gradual.


\subsection{Uncertainty Guided Geometric Regularization}

While the periodic TIDI cleanup acts as a reactive mechanism to remove floaters that have already formed, it is complemented by a proactive strategy designed to suppress their initial formation. This regularization is particularly critical in photometrically ambiguous regions, such as large textureless walls or glossy floors, where weak multi-view constraints make the optimization problem ill-posed. In these areas, the optimizer is prone to creating geometrically incorrect structures to satisfy the photometric loss. 

To address this, we incorporate a geometric prior derived from pre-trained monocular depth estimation networks, such as MiDaS-DPT \cite{MiDaS2022, DPT2021} or Depth Anything \cite{DepthAnything2024}. These networks provide a plausible, high-level understanding of the scene's structure. A critical challenge is that these priors are only accurate up to an unknown scale and shift and can contain significant errors, especially on challenging surfaces. Therefore, we use this prior not as a hard constraint, but as a form of soft guidance. 

During training, we introduce a loss term that minimizes the error between the rendered depth map ($\hat{Z}(p)$) and the monocular prediction ($\widetilde{Z}(p)$). To handle the inherent ambiguity, we solve for the optimal scale $(s)$ and shift $(t)$ parameters that align the two depth maps for each camera view. Crucially, the entire loss is guided by uncertainty to ensure robustness, following the principles of multi-task learning \cite{UncertaintyWeighting2018}. The uncertainty-aware depth loss is formulated as:

\begin{equation}
\mathcal{L}{\text{depth}} = \sum{p} w_{\text{uncert}}(p) \cdot \rho!\big(s,\hat{Z}(p)+t-\widetilde{Z}(p)\big)
\end{equation}

For each pixel $p$, the uncertainty weight $w_{\text{uncert}}(p)$ is a crucial term that reduces the loss contribution in regions where the depth estimator is intrinsically unreliable (e.g., on reflective surfaces, object boundaries, or transparent objects). This weighting scheme prevents the model from importing biases from the depth estimator and focuses the regularization on areas where the prior is most trustworthy. Furthermore, we use a robust Huber loss ($\rho$) \cite{HuberRobust1964}, which is less sensitive to large, occasional prediction errors than a standard L2 loss, making the training process more stable.

By applying this uncertainty-aware loss, we effectively ‘nudge’ the optimizer to form geometric structures that conform to the surfaces suggested by the depth prior. This discourages the placement of Gaussians in empty space, thereby suppressing the initial formation of floaters and reducing the burden on the subsequent cleanup stage. Finally, to ensure training stability, we introduce this depth loss gradually using a curriculum schedule. The influence of the depth term is increased only after the photometric reconstruction has become reasonably stable, preventing the geometric prior from overpowering the color information in the early stages of training. This approach provides robust geometric guidance without being overly restrictive, making it an effective complement to our evidence-based pruning, similar in spirit to depth supervision in NeRF \cite{DepthSupervisedNeRF2022}.

\section{Results and Evaluation}
\label{sec:results}
This section provides a comprehensive evaluation of our TIDI-GS. We begin by detailing the experimental setup and evaluation protocols designed to test the performance of our framework. We then present an in-depth analysis of our method's internal dynamics, followed by extensive qualitative and quantitative comparisons against several leading baseline methods on a variety of challenging indoor datasets. The results collectively demonstrate the effectiveness and efficiency of our floater suppression framework in producing high-fidelity 3D reconstructions. 

\begin{figure*}[t!]
    \centering
    \includegraphics[width=\textwidth]{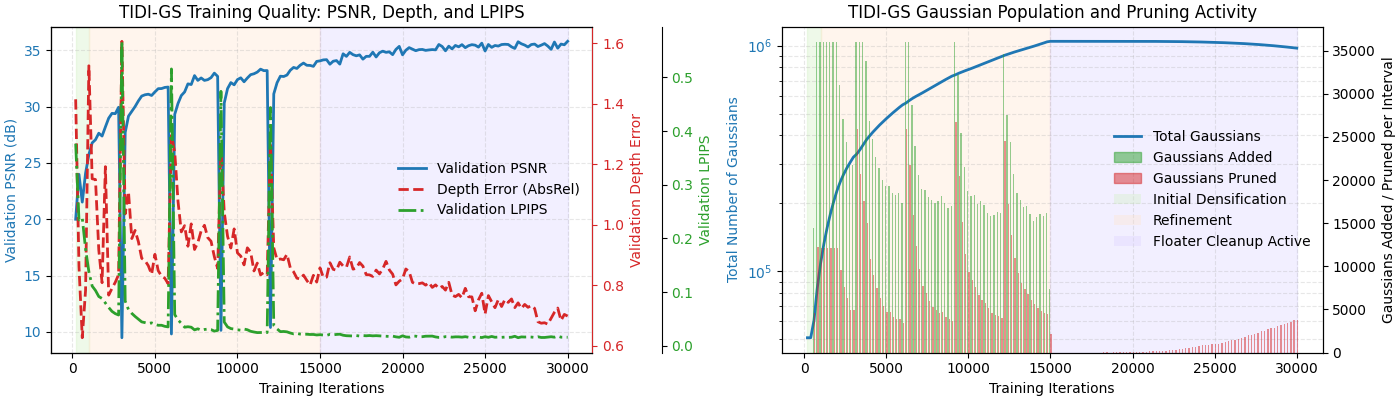}
    \caption{TIDI-GS training dynamics. (Left) Photometric (PSNR, LPIPS) and geometric (Depth Error) quality metrics improve concurrently throughout training. (Right) The total number of Gaussians grows and then stabilizes as the pruning activity (red bars) increases in the later stages, indicating a shift from a growth phase to a refinement phase. }
    \label{fig:evaluation_room}
\end{figure*}


\subsection{Implementation Details}

We implemented TIDI-GS in PyTorch to ensure a fair and reproducible comparison. It builds directly upon the publicly available 3DGS source code with its tile-based differentiable rasterizer \cite{3DGS2023}. This approach guarantees that the core rendering pipeline remains identical across all tested methods, and it isolates the performance differences to the contributions of our framework. All experiments were conducted on a single consumer-grade NVIDIA RTX 4080 GPU with 16 GB of memory. We employed mixed-precision rendering (FP16) for efficiency and memory savings, while maintaining full precision (FP32) for the optimizer state and EMA buffers to ensure numerical stability throughout the training process.

\begin{figure*}[h!]
  \centering
  \includegraphics[width=\textwidth]{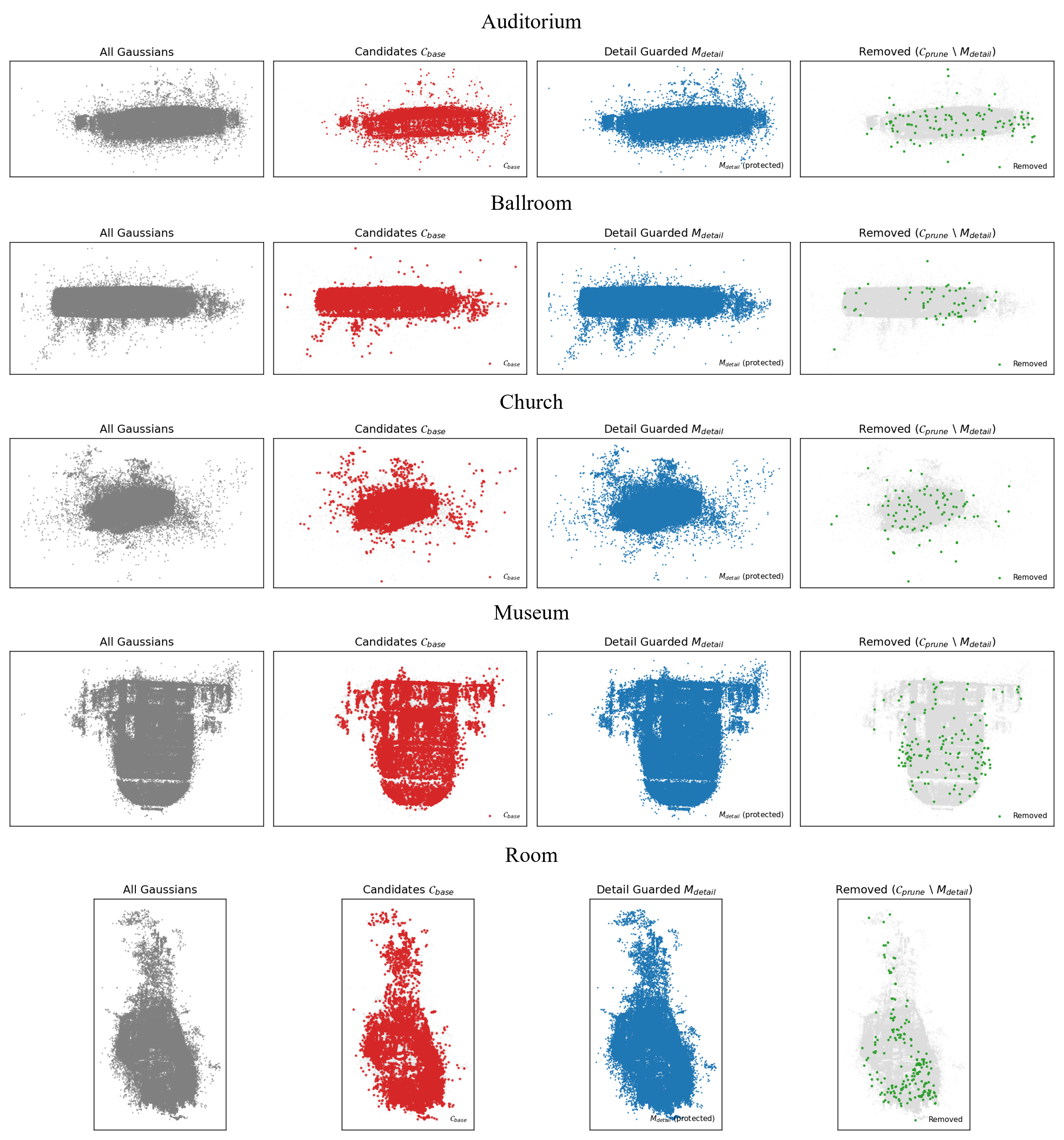}
  \caption{Visualization of the detail-aware pruning pipeline across different scenes. From left to right: (1) All Gaussians, (2) Candidate floaters ($\mathcal{C}_{\text{base}}$) in red, (3) Detail guarded candidates ($M_{\text{detail}}$) in blue, and (4) Finally removed Gaussians in green. The process effectively isolates sparse floaters while preserving the dense structure of the underlying geometry. }
  \label{fig:pruning}
\end{figure*}

Our evaluation is centered on five challenging indoor sequences: Auditorium, Ballroom, Church, and Museum from the Tanks and Temples dataset \cite{TanksnTemples2017}, and the Room scene from the Mip-NeRF 360 dataset \cite{MipNeRF3602022}. Each scene consists of 120-200 high-resolution (1600$\times$1200) images, which were pre-processed with COLMAP to obtain initial camera poses and a sparse point cloud. We trained each scene for a fixed 30,000 iterations, initializing the model with one anisotropic Gaussian per point from the COLMAP output. The standard 3DGS densification schedule was kept intact for all methods. Our \emph{TIDI} cleanup process was executed periodically every 400 training steps, following an initial 500-step period to allow the learnable importance parameters ($\omega$) to stabilize for newly spawned Gaussians. The photometric regularization was provided by a composite loss function combining L1, SSIM ($\lambda_{\text{SSIM}}=0.2$), and LPIPS (using AlexNet \cite{AlexNet2012} with a weight of 0.1), a combination chosen to effectively balance pixel-level accuracy with the preservation of structural and perceptual fidelity under complex indoor lighting \cite{LPIPS2018}. 

For the geometric regularization component, monocular depth priors were generated by fusing the outputs of MiDaS-DPT \cite{MiDaS2022, DPT2021} and Depth-Anything V2 \cite{DepthAnything2024, DepthAnythingV2_2024}. The fusion was weighted using precision estimates derived from flip consistency variance to prioritize more reliable depth predictions. The evidence ledger statistics, which include multi-view visibility and the position gradient EMA ($\beta=0.99$), were accumulated online with minimal overhead. The pruning process utilized a $k$-NN distance score (with $k$=16) and was constrained by adaptive local and global removal caps of 1.0\% and 0.2\% of the Gaussians per cell and per scene, respectively. All baseline methods, including the original 3DGS \cite{3DGS2023}, Pixel-GS \cite{PixelGS2024}, LP-GS \cite{LPGS2024}, and Micro-Splatting \cite{MicroSplatting2024}, were retrained from scratch on our data splits using their author-recommended hyperparameter settings. We found the method to be robust with respect to different pruning threshold choices. A detailed hyperparameter sensitivity analysis is provided in Table \ref{tab:hyperparameter}. 


\subsection{Evaluation Protocol}

The indoor scenes for our benchmark were specifically chosen because they contain a high concentration of features known to induce floater artifacts. These include high-intensity direct and indirect light sources, glossy and specular materials like polished floors and glass, large textureless regions such as walls and ceilings, and complex, repeated geometric patterns. To provide a thorough quantitative assessment, we report a suite of standard, full image quality metrics: Peak Signal-to-Noise Ratio (PSNR) for pixel-level fidelity, Structural Similarity Index (SSIM) for structural coherence, and Learned Perceptual Image Patch Similarity (LPIPS) \cite{LPIPS2018} for perceptual quality. 

\begin{figure*}[t!]
    \centering
    \includegraphics[width=\textwidth]{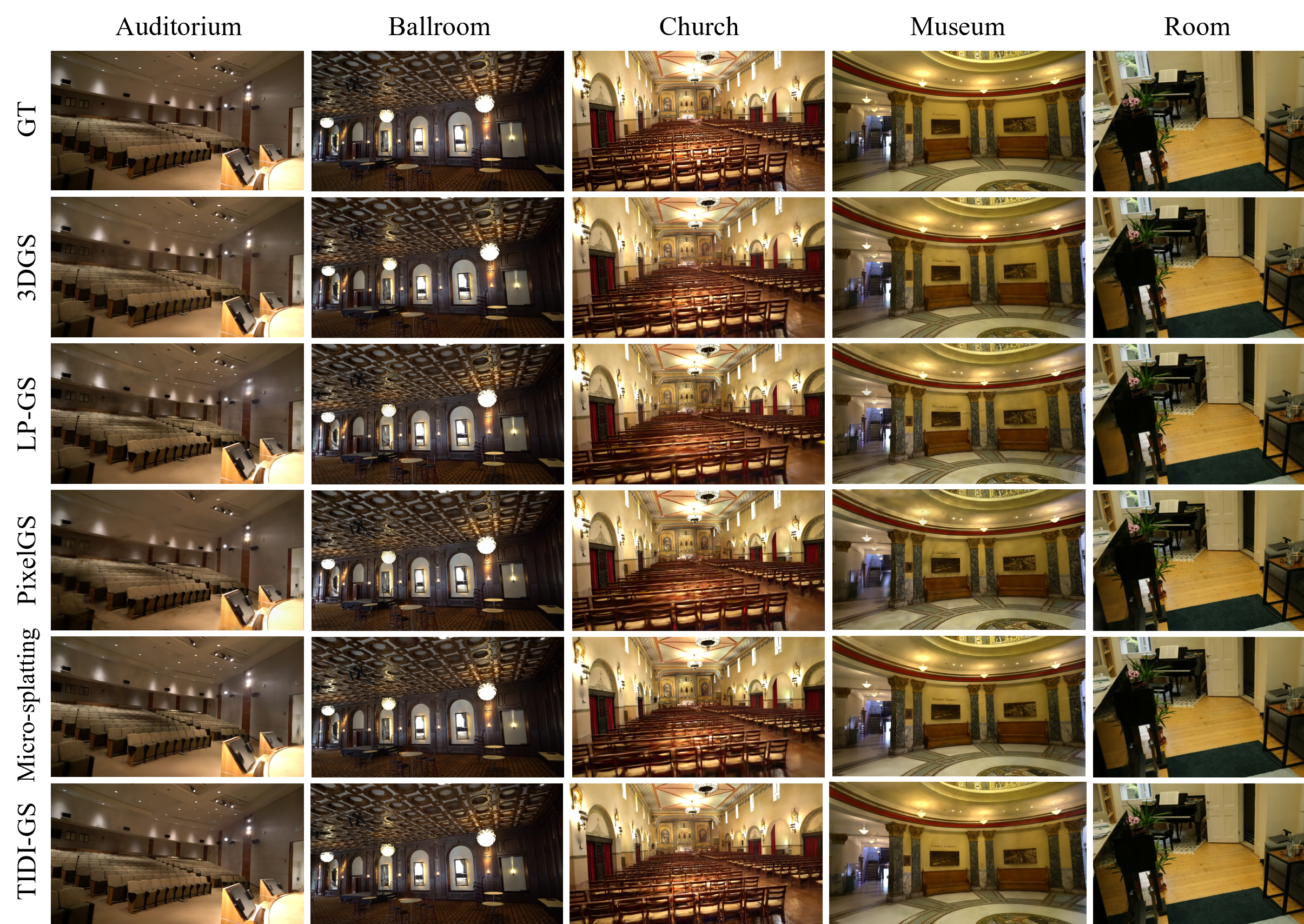}
    \caption{Qualitative comparison of rendering results on five indoor scenes at 30K training iterations. The columns correspond to scenes, and the rows correspond to different methods: Ground Truth (GT), baseline 3DGS, LP-GS, PixelGS, Micro-Splatting, and our TIDI-GS (bottom row). Our method consistently removes floater artifacts, haze, and blur present in baseline methods while preserving sharp details and clean surfaces, resulting in a reliable reconstruction compared to the ground truth. }
    \label{fig:compare}
\end{figure*}

\begin{figure*}[h!]
    \centering
    \includegraphics[width=0.8\textwidth]{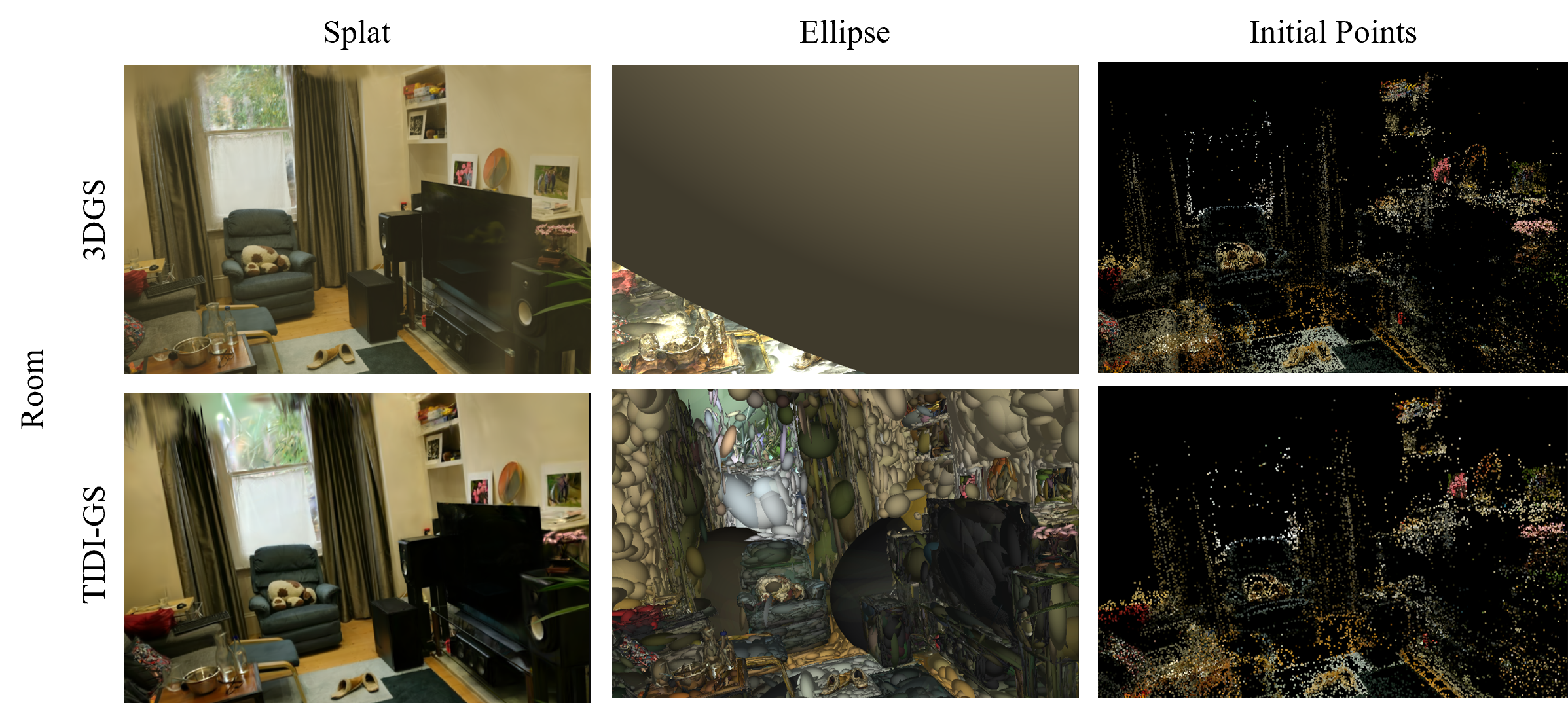}
    \caption{Qualitative comparison on challenging indoor scenes from the Mip-NeRF 360 dataset. From left to right, we present the input RGB images, geometry/appearance visualizations, and point-based representations. Baseline results exhibit noticeable geometric distortions and floating artifacts, while the improved results show cleaner geometry with more consistent structure and appearance across views.}
    \label{fig:sei_mipnerf}
\end{figure*}

\begin{figure*}[t!]
    \centering
    \includegraphics[width=0.82\textwidth]{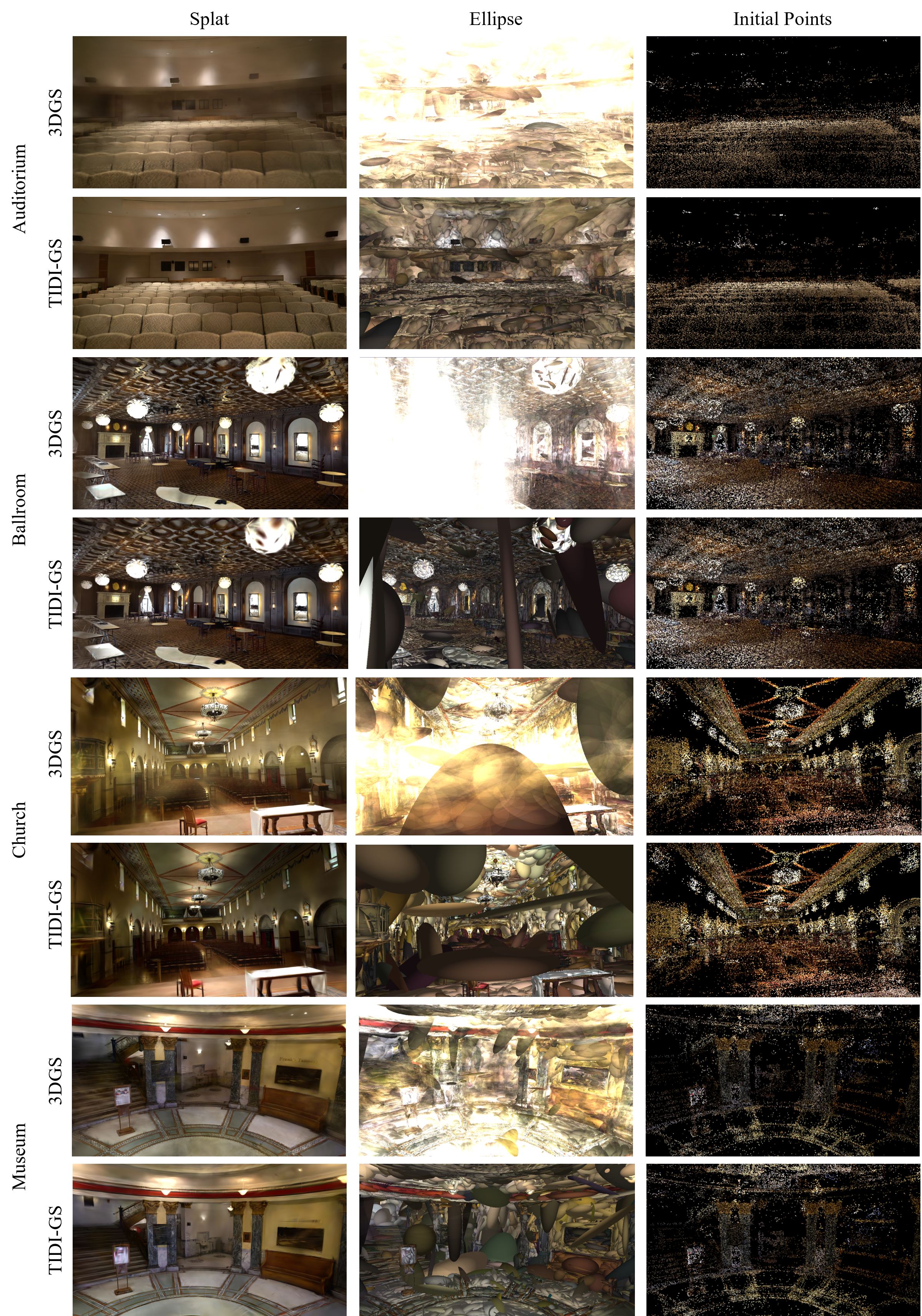}
    \caption{Qualitative results on large-scale indoor scenes from the Tanks and Temples benchmark. Each row compares different reconstruction stages, including rendered views, internal geometry visualizations, and point distributions. Baseline results exhibit heavy over-accumulation of Gaussians, interior clutter, and visibility leakage, particularly in texture-poor regions such as walls, ceilings, and long corridors. The improved results show substantially suppressed floating artifacts, enhanced spatial consistency, and clearer structural boundaries, especially in large halls and long-range indoor environments.}
    \label{fig:sei_tt}
\end{figure*}

However, it is well-known that these static, pixel-wise metrics often fail to capture the dynamic artifacts that most significantly degrade the user experience and reliability of interactively rendered 3D models \cite{ZipNeRF2023, RefNeRF2022}. Artifacts such as temporal shimmering of translucent surfaces, light leaks across silhouette boundaries, and sudden spikes in the depth map during camera motion are poorly penalized by metrics that average errors over the entire image. Therefore, we complement these standard scores with a targeted analysis of cleanliness indicators. These are computed on short or rendered video sequences with small camera jitters around validation viewpoints and include: (i) silhouette leakage: quantifies the fraction of foreground-mask pixels whose rendered transmittance incorrectly exceeds a small threshold outside of the ground truth object contours, (ii) depth stability: measures the proportion of pixels whose rendered depth remains stable (below a fixed tolerance) under small camera noise, and (iii) background consistency: measures the PSNR of static scene regions that should remain unchanged. By pairing conventional metrics with this stability-focused analysis, we provide a more holistic and practically relevant assessment of reconstruction quality.


\subsection{Analysis of TIDI-GS Behavior}

To verify that the components of TIDI-GS function as intended, we first analyze its internal training dynamics. Fig. \ref{fig:evaluation_room} provides a detailed look at the evolution of key metrics throughout the training process. The plot on the left demonstrates that as training progresses, the validation PSNR (blue) steadily increases while the depth error (red) and perceptual LPIPS metric (green) consistently decrease. This concurrent improvement indicates a healthy optimization process where photometric quality and geometric accuracy are refined in parallel. The plot on the right details the management of the Gaussian population. The total count (blue line) grows rapidly during the initial densification phase, driven by the standard 3DGS logic. It then stabilizes and begins to slightly decrease once our periodic \emph{TIDI} cleanup process activates. The bar chart shows that while new Gaussians are added throughout training (green bars), the pruning activity (red bars) becomes more significant and consistent in the later stages. This illustrates a shift from a growth-dominated phase to a refinement-dominated phase, where our method efficiently removes redundant or unsupported primitives without causing instability. 

The pruning pathway itself is visualized in greater detail in Fig. \ref{fig:pruning}. For a representative scene, we start with the entire set of Gaussians (leftmost column). Our method first identifies a broad, over-inclusive set of candidates (second column, red) that exhibit weak evidence across multiple criteria. The third column powerfully illustrates the crucial role of the detail-preserving guards: it shows in blue the subset of candidates that were protected from pruning because they were identified as being part of consistent or high-frequency structures. The final set of removed Gaussians (rightmost column, green) consists almost entirely of sparse, isolated points located in free space. This visual breakdown confirms that our method successfully isolates and excises floaters while carefully preserving the dense geometry of the underlying scene. 

\begin{figure}[h!]
  \centering
  \includegraphics[width=\linewidth]{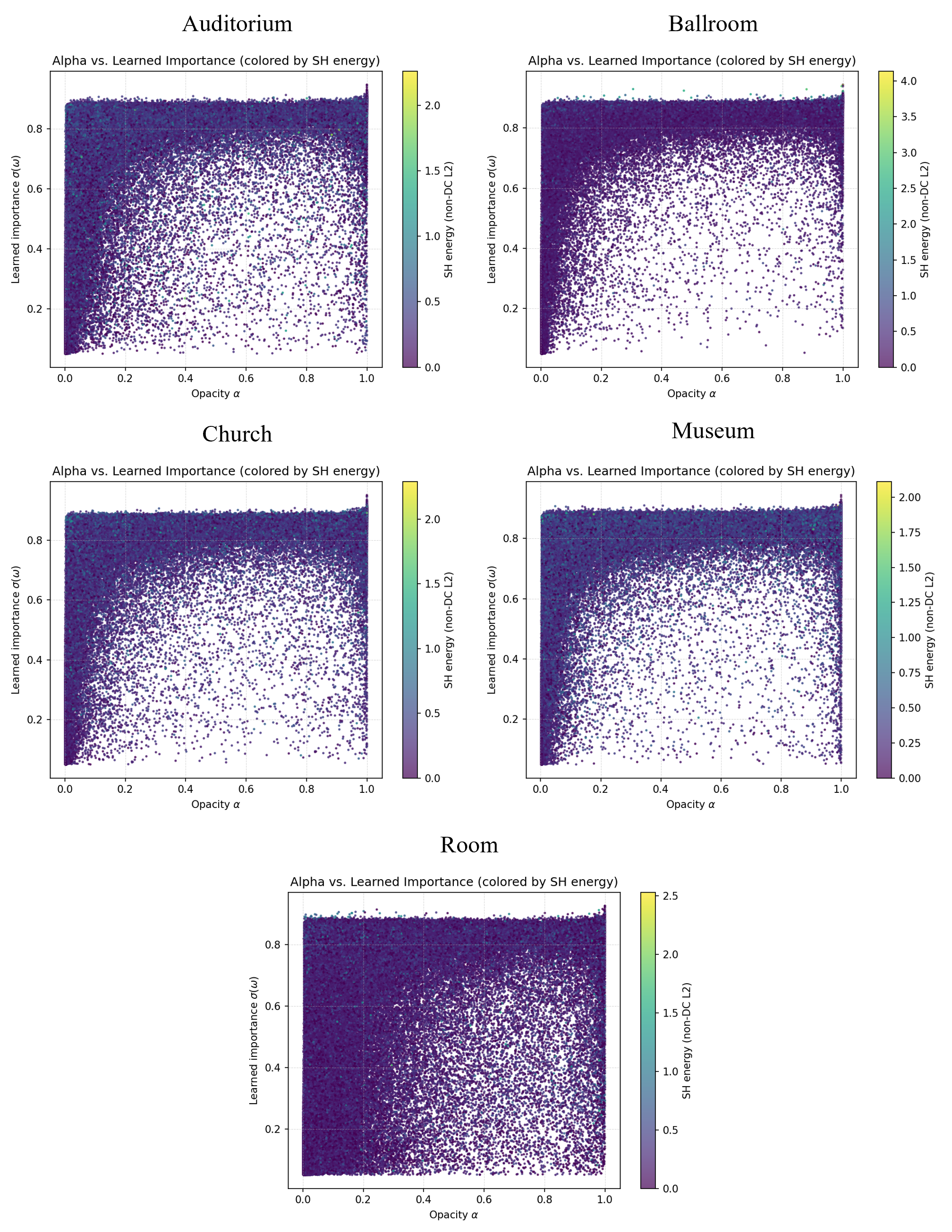}
  \caption{Analysis of Gaussian properties across the five test scenes. The plots show many low-opacity Gaussians with high spherical harmonics (SH) energy, which correspond to important specular details. This indicates that opacity alone is not sufficient for pruning and motivates the use of our detail-preserving guards. }
  \label{fig:alpha_omega}
\end{figure}

The learned behavior of our framework is further evidenced in Fig. \ref{fig:alpha_omega} which plots the relationship between opacity and learned importance, with each Gaussian colored by its non-DC SH energy. A critical insight from this analysis is the significant population of yellow points, representing high SH energy that persist even at very low opacity levels. These points correspond to specular details on surfaces which contribute significantly to realism. A naive opacity-based pruner would likely eliminate these valuable primitives. Our detail-aware guards, however, correctly identify their high SH energy and protect them, highlighting the sophistication of our method beyond simple thresholding. The plots also show a dense concentration of low-importance, low-opacity points that are the primary targets of our pruning policy. 

\subsection{Qualitative Comparison}
The most direct evidence of our method’s effectiveness is found in a qualitative comparison of the final rendered images. Fig. \ref{fig:compare} presents a side-by-side evaluation of TIDI-GS against several leading baseline methods across five challenging indoor test scenes. A consistent pattern emerges: our method produces reconstructions that are noticeably cleaner and more detailed than the baselines. Common artifacts from the other methods, such as semi-transparent veils, hazy glows around lights, and distinct free-space floaters, are effectively eliminated by TIDI-GS. As a result, the renders produced by TIDI-GS not only exhibit higher perceptual quality, but also align closely with the ground truth.

To understand the source of these visual improvements, Fig. \ref{fig:sei_mipnerf} and Fig. \ref{fig:sei_tt} offer a diagnostic look into the underlying geometric representation learned by our method in comparison to the baseline 3DGS. These figures deconstruct the reconstruction of two representative scenes into the following components: the final rendered image (‘Splat’), a visualization of the 3D Gaussians as geometric ellipses (‘Ellipse’), and the sparse ‘Initial Points’ from which both methods began training. This comparison makes it clear that TIDI-GS’s superior results stem from its ability to learn a fundamentally cleaner and more geometrically accurate set of primitives, rather than from any superficial post-processing effect. 

\begin{table*}[t!]
    \centering
    \caption{Quantitative result of Tanks and Temples \cite{TanksnTemples2017} and Mip-NeRF 360 \cite{MipNeRF3602022} (30K iterations)}
    \label{tab:comparison}
    \definecolor{first}{RGB}{255, 179, 179}
    \definecolor{second}{RGB}{255, 223, 179}
    \definecolor{third}{RGB}{255, 255, 204}
    \begin{minipage}[t]{0.45\textwidth}
        \centering
        \resizebox{\linewidth}{!}{
        \begin{tabular}{l|cccc}
        \multicolumn{5}{c}{Auditorium}\\
        & SSIM$\uparrow$ & PSNR$\uparrow$ & LPIPS$\downarrow$ & Time$\downarrow$ \\
        \midrule
             3DGS &\cellcolor{first} 0.935 & \cellcolor{first} 30.8 & \cellcolor{third} 0.109 & \cellcolor{third} 24 min \\
             LP-GS & \cellcolor{third} 0.906 & 29.0 & 0.257 & 31 min\\
             Micro-Splatting & 0.852 & \cellcolor{second} 29.8 & \cellcolor{second} 0.106 & \cellcolor{first} 9 min\\
             PixelGS & 0.830 & 24.1 & 0.351 & 28 min\\
             TIDI-GS & \cellcolor{second} 0.934 & \cellcolor{third} 29.6 & \cellcolor{first} 0.028 & \cellcolor{second} 20 min\\
        \end{tabular}
        }
    \end{minipage}
    \vspace{0.2cm}
    \begin{minipage}[t]{0.45\textwidth}
        \centering
        \resizebox{\linewidth}{!}{
        \begin{tabular}{l|cccc}
        \multicolumn{5}{c}{Ballroom}\\
        & SSIM$\uparrow$ & PSNR$\uparrow$ & LPIPS$\downarrow$ & Time$\downarrow$ \\
        \midrule
             3DGS & \cellcolor{first} 0.933 & \cellcolor{second} 28.5 & \cellcolor{second} 0.050 & 27 min \\
             LP-GS & 0.867 & 26.2 & 0.152 & 45 min\\
             Micro-Splatting & \cellcolor{second} 0.9254 & \cellcolor{third} 26.3 & \cellcolor{third} 0.073 & \cellcolor{first} 15 min\\
             PixelGS & 0.818 & 24.4 & 0.191 & 22 min\\
             TIDI-GS & \cellcolor{third} 0.918 & \cellcolor{first} 28.8 & \cellcolor{first} 0.023 & \cellcolor{second} 19 min\\
        \end{tabular}
        }
    \end{minipage}
    \vspace{0.2 cm}
    \begin{minipage}[t]{0.45\linewidth}
        \centering
        \resizebox{\linewidth}{!}{
        \begin{tabular}{l|cccc}
        \multicolumn{5}{c}{Church}\\
        & SSIM$\uparrow$ & PSNR$\uparrow$ & LPIPS$\downarrow$ & Time$\downarrow$ \\
        \midrule
             3DGS & \cellcolor{first} 0.898 & \cellcolor{first} 26.2 & \cellcolor{third} 0.128 & \cellcolor{second} 21 min\\
             LP-GS & \cellcolor{third} 0.825 & 24.3 & 0.269 & 40 min\\
             Micro-Splatting & 0.797 & 24.6 & \cellcolor{second} 0.117 & \cellcolor{first} 17 min\\
             PixelGS & 0.781 & \cellcolor{third} 24.7 & 0.287 & \cellcolor{third} 33 min\\
             TIDI-GS & \cellcolor{second} 0.865 & \cellcolor{second} 25.4 & \cellcolor{first} 0.049 & \cellcolor{second} 21 min\\
        \end{tabular}
        }
    \end{minipage}
    \vspace{0.2 cm}
    \begin{minipage}[t]{0.45\linewidth}
        \centering
        \resizebox{\linewidth}{!}{
        \begin{tabular}{l|cccc}
        \multicolumn{5}{c}{Museum}\\
        & SSIM$\uparrow$ & PSNR$\uparrow$ & LPIPS$\downarrow$ & Time$\downarrow$ \\
        \midrule
             3DGS & \cellcolor{first} 0.914 & \cellcolor{first} 27.3 & \cellcolor{third} 0.127 & 29 min\\
             LP-GS & 0.854 & \cellcolor{third} 25.8 & 0.196 & 52 min\\
             Micro-Splatting & \cellcolor{third} 0.868 & 25.7 & \cellcolor{second} 0.089 & \cellcolor{first} 17 min\\
             PixelGS & 0.781 & 22.3 & 0.255 & \cellcolor{second} 21 min\\
             TIDI-GS & \cellcolor{second} 0.904 & \cellcolor{second} 26.7 & \cellcolor{first} 0.028 & \cellcolor{third} 23 min\\
        \end{tabular}
        }
    \end{minipage}
    \vspace{0.2 cm}
    \begin{minipage}[t]{0.45\linewidth}
        \centering
        \resizebox{\linewidth}{!}{
        \begin{tabular}{l|cccc}
        \multicolumn{5}{c}{Room}\\
        & SSIM$\uparrow$ & PSNR$\uparrow$ & LPIPS$\downarrow$ & Time$\downarrow$ \\
        \midrule
             3DGS & \cellcolor{second} 0.959 & \cellcolor{first} 35.9 & \cellcolor{third} 0.079 & \cellcolor{third} 23 min \\
             LP-GS & \cellcolor{third} 0.951 & \cellcolor{second} 34.6 & 0.0976 & 25 min\\
             Micro-Splatting & 0.922 & \cellcolor{second} 24.6 & \cellcolor{second} 0.058 & \cellcolor{first} 8 min \\
             PixelGS & 0.933 & \cellcolor{third} 31.9 & 0.180 & 27 min \\
             TIDI-GS& \cellcolor{first} 0.964 & \cellcolor{first} 35.9 & \cellcolor{first} 0.016 & \cellcolor{second} 16 min
        \end{tabular}
        }
    \end{minipage}
\end{table*}

\begin{table*}[h]
\centering
\caption{Sensitivity analysis for key pruning thresholds on the Room scene. We vary the visibility count threshold ($\tau_{\text{vis}}$) and the gradient EMA threshold ($\tau_{\text{grad}}$). The results show that our method is robust, with optimal performance achieved at $\tau_{\text{vis}}=2.0$ and $\tau_{\text{grad}}=5.00E-04$. }
\label{tab:hyperparameter}
\definecolor{first}{RGB}{255, 179, 179}
\definecolor{second}{RGB}{255, 223, 179}
\begin{minipage}[h]{0.45\textwidth}
\centering
\resizebox{\linewidth}{!}{
    \begin{tabular}{l|cccccc}
    \multicolumn{7}{c}{Visibility ($\tau_{\text{vis}}$)}\\
    Metrics & SSIM$\uparrow$ & PSNR$\uparrow$ & LPIPS$\downarrow$ & Loss$\downarrow$ & AbsRel$\downarrow$ & $\delta$1.25$\uparrow$ \\
    \hline
    1.0 & \cellcolor{second} 0.9518 & 33.9525 & \cellcolor{second} 0.0234 & 0.0136 & 0.742 & 0.621 \\
    2.0 & \cellcolor{first} 0.9643 &  \cellcolor{first} 35.8983 & \cellcolor{first} 0.016 & \cellcolor{first} 0.0104 & \cellcolor{first} 0.6979 & \cellcolor{first} 0.6597 \\
    3.0 & 0.951 & \cellcolor{second} 33.9598 & 0.0238 & \cellcolor{second} 0.01356 & \cellcolor{second} 0.7135 & \cellcolor{second} 0.639 \\
    \end{tabular}
    }
\end{minipage}
\begin{minipage}[h]{0.45\textwidth}
\centering
\resizebox{\linewidth}{!}{
    \begin{tabular}{l|cccccc}
    \multicolumn{7}{c}{Grad ($\tau_{\text{grad}}$)}\\
    Metrics & SSIM$\uparrow$ & PSNR$\uparrow$ & LPIPS$\downarrow$ & Loss$\downarrow$ & AbsRel$\downarrow$ & $\delta$1.25$\uparrow$ \\
    \hline
    1.00E-04 & 0.9515 & 33.9543 & 0.0237 & 0.0135 & 0.735 & 0.6305 \\
    5.00E-04 & \cellcolor{first} 0.9643 & \cellcolor{first} 35.8983 & \cellcolor{first} 0.016 & \cellcolor{first} 0.0104 & \cellcolor{first} 0.6979 & \cellcolor{first} 0.6597 \\
    1.00E-03 & \cellcolor{second} 0.9517 & \cellcolor{second} 34.0208 & \cellcolor{second} 0.0235 & \cellcolor{second} 0.0134 & \cellcolor{second} 0.709 & \cellcolor{second} 0.642\\
    \end{tabular}
    }
\end{minipage}
\begin{minipage}[h]{0.45\textwidth}
\centering
\resizebox{\linewidth}{!}{
    \begin{tabular}{l|cccccc}
    \multicolumn{7}{c}{Opacity ($\tau_\alpha$)}\\
    Metrics & SSIM$\uparrow$ & PSNR$\uparrow$ & LPIPS$\downarrow$ & Loss$\downarrow$ & AbsRel$\downarrow$ & $\delta$1.25$\uparrow$ \\
    \hline
    0.01 & \cellcolor{second} 0.9515 & \cellcolor{second} 34.023 & 0.0238 & \cellcolor{second} 0.0133 & 0.752 & 0.618 \\
    0.04 & \cellcolor{first} 0.9643 & \cellcolor{first} 35.8983 & \cellcolor{first} 0.016 & \cellcolor{first} 0.0104 & \cellcolor{first} 0.6979 & \cellcolor{first} 0.6597 \\
    0.08 & \cellcolor{second} 0.9515 & 33.9695 & \cellcolor{second} 0.0236 & 0.0135 & \cellcolor{second} 0.7210 & \cellcolor{second} 0.6360 \\
    \end{tabular}
    }
\end{minipage}
\begin{minipage}[h]{0.45\textwidth}
\centering
\resizebox{\linewidth}{!}{
    \begin{tabular}{l|cccccc}
    \multicolumn{7}{c}{Importance ($\tau_\omega$)}\\
    Metrics & SSIM$\uparrow$ & PSNR$\uparrow$ & LPIPS$\downarrow$ & Loss$\downarrow$ & AbsRel$\downarrow$ & $\delta$1.25$\uparrow$ \\
    \hline
    0.2 & \cellcolor{second} 0.9521 & \cellcolor{second} 34.0135 & \cellcolor{second} 0.0236 & \cellcolor{second} 0.0134 & \cellcolor{second} 0.7240 & \cellcolor{second} 0.6420 \\
    0.35 & \cellcolor{first} 0.9643 & \cellcolor{first} 35.8983 & \cellcolor{first} 0.016 & \cellcolor{first} 0.0104 & \cellcolor{first} 0.6979 & \cellcolor{first} 0.6597 \\
    0.5 & 0.9513 & 33.9578 & 0.0237 & \cellcolor{second} 0.0134 & 0.7310 & 0.6330 \\
    \end{tabular}
    }
\end{minipage}
\end{table*}

This difference is clearly demonstrated in the Auditorium scene: the baseline 3DGS render is obscured by a pervasive haze that reduces contrast and washes out details. The corresponding ‘Ellipse’ visualization reveals the cause, a chaotic, noisy cloud of poorly structured primitives. In contrast, the TIDI-GS render is sharp and clear throughout. Its ‘Ellipse’ view shows a clean, well-organized set of Gaussians that conform closely to the actual surfaces of the scene, indicating a far more coherent underlying geometric structure.

While the Auditorium example shows significant improvement, the difference is even more striking in the Room scene, where the baseline method suffers a catastrophic failure. It generates a large, scene-occluding floater that completely obstructs the view. As shown in Fig. \ref{fig:sei_mipnerf}, TIDI-GS completely eliminates this artifact. The ‘Ellipse’ view reveals why: the baseline has populated what should be empty space with a dense, unstructured cluster of large, unsupported Gaussians. TIDI-GS, guided by its evidence-based pruning and geometric regularization, completely avoids this pathological behavior. Crucially, both methods started from the same initial points, proving that this dramatic improvement is a direct result of our proposed training framework.

This principle of improved geometric organization explains TIDI-GS’s superior results across all test cases in Fig. \ref{fig:compare}. For example, TIDI-GS renders the intricate patterns of the Ballroom ceiling and the fine wooden details of the Church chairs with crisp definition, free from the hazy artifacts that degrade the other reconstructions. In all instances, the visual evidence shows that TIDI-GS produces reconstructions that are not only cleaner on the surface, but also more robust and faithful in their underlying geometric representation.


\subsection{Quantitative Comparison}

The quantitative results, presented in Table \ref{tab:comparison}, provide strong statistical support for the visual trends observed in our qualitative analysis. Across the five challenging indoor scenes, TIDI-GS achieves PSNR and SSIM scores that are highly competitive with, and in several cases exceed, the baseline 3DGS method. This demonstrates that our framework successfully preserves the high-fidelity photometric accuracy that is a hallmark of Gaussian Splatting. The most significant and consistent advantage, however, is observed in the LPIPS metric, where TIDI-GS delivers the lowest (\emph{best}) scores across all scenes, often by a substantial margin. This indicates a major improvement in perceptual quality, as the LPIPS metric is designed to be more aligned with human perception of detail, texture, and structural correctness. 

Beyond raw image quality, the efficiency of our method is a critical aspect of its design. The quality gains are achieved with a training time that remains comparable to the baseline 3DGS, confirming the computational efficiency of our lightweight, plug-in approach. The final model sizes (measured by the number of Gaussians) are also similar to the baseline, which is a crucial finding. It demonstrates that TIDI-GS improves quality not by naively increasing model capacity, but by intelligently optimizing the existing primitives and, most importantly, by removing false or unnecessary elements. This leads to a more compact and efficient representation of the scene geometry. 

While standard metrics offer a static snapshot, our evaluation also focuses on dynamic ‘cleanliness’ indicators that directly quantify the distracting artifacts that appear during interactive viewing. To quantify the improvements in geometric integrity that are not fully captured by these metrics, we evaluate silhouette leakage and depth stability under minor camera motion. The results, presented in the ablation study in Table \ref{tab:ablation}, show a clear advantage for the full TIDI-GS model. Our method significantly reduces silhouette leakage, which indicates cleaner and more accurate object boundaries without the hazy artifacts common in other methods. Furthermore, it achieves the highest depth stability, confirming that our geometrically regularized reconstructions are more temporally consistent and less prone to the distracting shimmering or ‘popping’ artifacts. This demonstrates the practical benefit of our floater suppression framework in creating robust and reliable digital assets. 

Finally, a practical method must not only perform well but also be robust to its hyperparameter settings, avoiding the need for extensive, per-scene tuning. To validate the robustness of our \emph{TIDI} cleanup framework, we conducted a sensitivity analysis on all four of its key pruning thresholds: the visibility count ($\tau_{\text{vis}}$), the position gradient EMA ($\tau_{\text{grad}}$), opacity ($\tau_\alpha$), and learned importance ($\tau_\omega$). The comprehensive results are shown in Table \ref{tab:hyperparameter}. Across all four experiments, a clear trend emerges: there is a ‘sweet spot’ where the thresholds are strict enough to effectively remove floaters but not so aggressive that they begin to damage legitimate scene content. For example, a visibility threshold ($\tau_{\text{vis}}$) of 2.0 achieves the best balance across all metrics. Setting the threshold too low (1.0) or too high (3.0) leads to a noticeable degradation in performance. This same pattern holds for the other parameters, with optimal performances found at $\tau_{\text{grad}}=5.00E-04$, $\tau_\alpha=0.04$, and $\tau_\omega=0.35$. This analysis demonstrates that while the choice of thresholds is important, our method is not overly sensitive and performs robustly with a single, well-chosen set of parameters across different scenes, underscoring its practicality. 


\subsection{Ablation Study}

\begin{figure*}[t!]
    \centering
    \includegraphics[width=\textwidth]{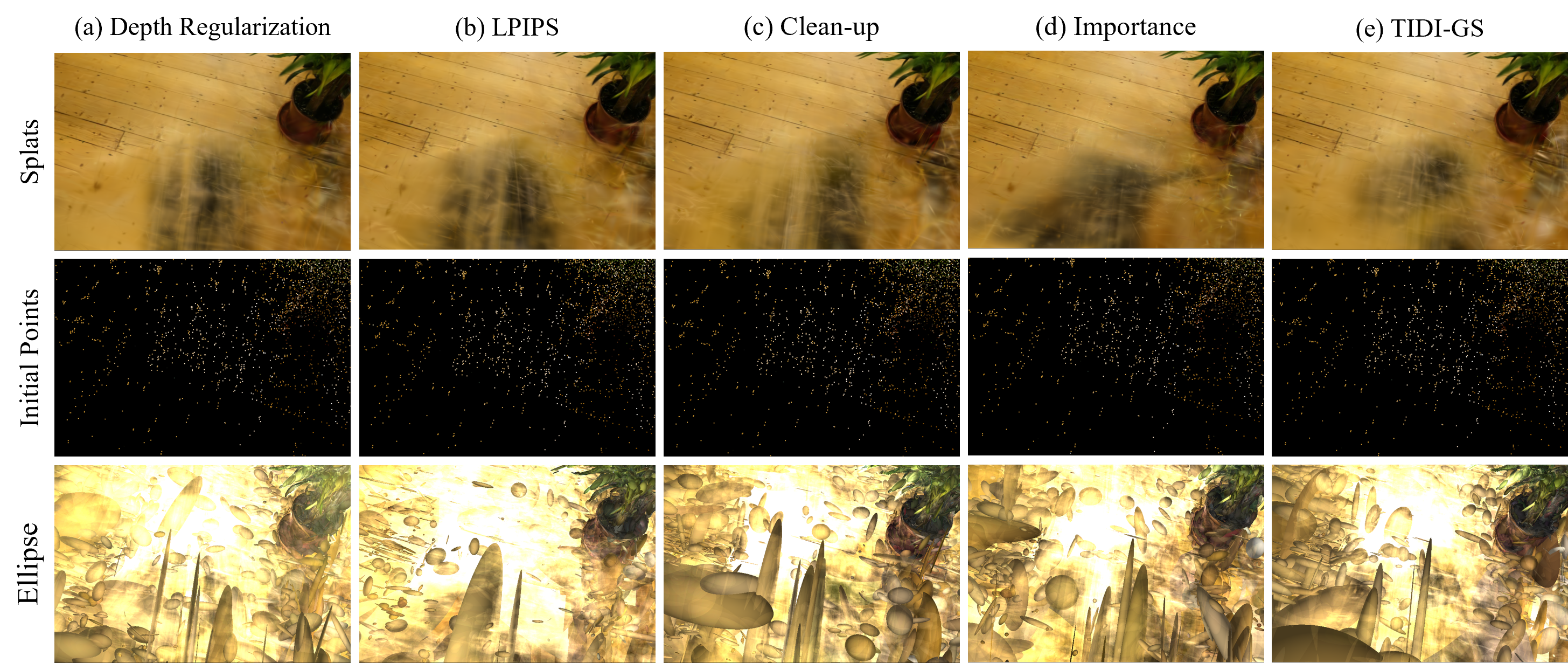}
    \caption{Qualitative ablation study on an indoor scene. The full TIDI-GS model (Left) is compared to representative ablations. Disabling key components like floater cleanup, depth regularization, LPIPS terms, and importance regularization reintroduces artifacts like translucent veils and wavy geometry, demonstrating the necessity of each component for both the final render and the underlying splat structure. }
    \label{fig:ablation}
\end{figure*}

To validate the design of TIDI-GS and isolate the contribution of its key components, we conducted a comprehensive ablation study. We disabled individual mechanisms while keeping the training budget and camera paths identical across all runs. The following analysis focuses on five of the most critical factors that highlight our framework, (a) \textbf{depth regularization}: the uncertainty guided monocular depth priors, (b) \textbf{LPIPS term}: the perceptual guidance provided by the LPIPS loss term, (c) \textbf{floater cleanup}: the core of our reactive artifact removal system, (d) \textbf{importance regularization}: the learned and data-driven aspect of our evidence log, and (e) \textbf{full TIDI-GS}. Qualitative comparisons are shown in Fig. \ref{fig:ablation}, with the pruning pipeline visualized in Fig. \ref{fig:pruning}, opacity-importance distributions in Fig. \ref{fig:alpha_omega}, scene comparisons in Fig. \ref{fig:compare}, and full numeric results in Table \ref{tab:ablation}. 

\begin{enumerate}[leftmargin=*, label=(\alph*)]
    \item Depth regularization: it provides a geometric foundation for the reconstruction, particularly for large, textureless surfaces where photometric evidence from multi-view images is weak or ambiguous. It acts as a proactive regularizer, and it encourages the formation of plausible and smooth surfaces from the early stages of training. 
    \item LPIPS Term: while standard L1 and L2 losses ensure pixel-level accuracy, the LPIPS term encourages the model to reproduce features that are important to human perception, such as sharp edges, fine textures, and complex structural details, which is critical for high-fidelity rendering. 
    \item Floater Cleanup: it encompasses the entire three-step \emph{TIDI}: identifying a broad set of candidates based on the evidence log, applying detail-preserving guards, and performing isolation-aware removal of the remaining unguarded primitives. Its purpose is to surgically excise the residual, high-frequency artifacts (floaters) that persist even with a stable geometric base. 
    \item Importance Regularization: by associating a learnable scalar importance weight, $\omega$, with each Gaussian, we allow the model to determine which primitives are most useful for reconstruction by moving beyond purely hand-crafted heuristics like opacity or visibility. 
\end{enumerate}

\begin{table*}[h]
\centering
\caption{Ablation study: performance across different metrics and configurations. Configurations (a)-(e) correspond to the ablation settings described in Section V-E.}
\label{tab:ablation}
\begin{minipage}[h]{0.45\textwidth}
\centering
\resizebox{\linewidth}{!}{
    \begin{tabular}{l|ccccc}
    \textbf{Metrics} & \textbf{(a)} & \textbf{(b)} & \textbf{(c)} & \textbf{(d)} & \textbf{(e) Ours} \\
    \hline
    SSIM$\uparrow$ & 0.9637 & 0.9511 & 0.9646 & 0.9645 & 0.9643 \\
    PSNR$\uparrow$ & 35.7709 & 34.5192 & 35.9617 & 35.8903 & 35.8983 \\
    LPIPS$\downarrow$ & 0.0165 & 0.0123 & 0.0160 & 0.0161 & 0.0100 \\
    Loss$\downarrow$ & 0.0229 & 0.0299 & 0.0205 & 0.0150 & 0.0104 \\
    Time$\downarrow$ & 24 min & 27 min & 35 min & 21 min & 16 min \\
    Size$\downarrow$ & 245.6 MB & 283.1 MB & 240.7 MB & 265.2 MB & 241.4 MB \\
    \end{tabular}
    }
\end{minipage}
\begin{minipage}[h]{0.45\textwidth}
\centering
\resizebox{\linewidth}{!}{
    \begin{tabular}{l|ccccc}
    \textbf{Metrics} & \textbf{(a)} & \textbf{(b)} & \textbf{(c)} & \textbf{(d)} & \textbf{(e) Ours} \\
    \hline
    AbsRel$\downarrow$ & 0.6283 & 0.9033 & 0.7373 & 0.7395 & 0.6979 \\
    $\delta_{1.25}\uparrow$ & 0.7142 & 0.6353 & 0.7192 & 0.6699 & 0.6597 \\
    Added & 1,682,010 & 2,563,112 & 1,672,244 & 1,661,167 & 1,678,696 \\
    Pruned & 64,929 & 1,472,282 & 752,540 & 643,269 & 755,996 \\
    Gaussians & 990,145 & 1,141,636 & 970,510 & 1,069,215 & 973,506 \\
    \end{tabular}
    }
\end{minipage}
\end{table*}

These ablations trace a clear causal chain. When the cleanup stack (c) is disabled, unsupported translucent splats persist and dominate the long-tail failures like shimmer and leakage. Removing depth regularization (a) specifically destabilizes large low-texture spans, and it seeds the halos that later manifest as jitter spikes. Dropping the LPIPS term (b) flattens the perceptual structure, causing edges to fray and textures to blur, which raises leakage even when PSNR/SSIM remain similar. Eliminating importance regularization (d) inflates the model with low utility splats, as seen in the opacity-importance scatter in Fig. \ref{fig:alpha_omega} and the heavier model footprint in Table \ref{tab:ablation}. In contrast, the full TIDI-GS model balances these forces: evidence-aware pruning removes truly isolated, low usage splats, detail guards retain high frequency content, and uncertainty-guided depth regularization suppresses nascent drift, yielding the robust and high fidelity reconstructions seen in Fig. \ref{fig:compare}.

\section{Discussion and Conclusion}
We presented TIDI-GS, a training framework that addresses a central failure mode of 3D Gaussian Splatting in indoor environments: unsupported, semi-transparent floaters that degrade both visual quality and measurement fidelity. Our method combines a temporally accumulated evidence log with conservative detail-preserving guards, an isolation-aware removal policy, and an uncertainty-guided monocular depth regularizer to effectively reduce these artifacts without requiring architectural changes or significant computational overhead. The success of this approach is rooted in its explicit optimization for indoor reconstructions, where the geometry is at a finite depth and richly constrained by multi-view overlaps. Our evidence-based pruning is highly effective because the concept of geometric support is well-defined, and the monocular depth prior is most reliable on the large, low-texture surfaces common indoors, allowing it to guide geometry without overregularizing complex, high-frequency details. This specialization for indoor scenes, however, also defines the framework's primary limitation: scenes with unbounded depth, such as outdoor environments with distant horizons and sky domes, where our assumptions about evidence and depth reliability become less stable. We view this not as a flaw but as a deliberate scope choice, which in turn points toward straightforward future extensions such as incorporating dedicated background models to broaden the method's applicability without sacrificing the simplicity of the current design. Other than improving rendering aesthetics, the practical outcome of TIDI-GS is the creation of more stable and geometrically reliable reconstructions. By aligning the 3DGS training loop with multi-view evidence and geometric uncertainty, our work advances the methodology toward the generation of robust digital twins for indoor analysis and offers a principled and efficient template for future artifact-aware optimization in splat-based rendering. 

\section*{Acknowledgments}
This work was supported by the National Research Foundation of Korea (NRF) grant funded by the Korea government (MSIT) (Grant No. RS-2022-NR067080 and RS-2025-05515607).


\bibliographystyle{IEEEtran}
\nocite{*}
\bibliography{ref}

@inproceedings{MipNeRF2021,
  author    = {Barron, Jonathan T. and Mildenhall, Ben and Tancik, Matthew and Hedman, Peter and Martin-Brualla, Ricardo and Srinivasan, Pratul P.},
  title     = {Mip-NeRF: A Multiscale Representation for Anti-Aliasing Neural Radiance Fields},
  booktitle = {Adv. Neural Inf. Process. Syst. (NeurIPS)},
  year      = {2021}
}

@inproceedings{MipNeRF3602022,
  author    = {Barron, Jonathan T. and Mildenhall, Ben and Verbin, Daniel and Srinivasan, Pratul P. and Hedman, Peter},
  title     = {Mip-NeRF 360: Unbounded Anti-Aliased Neural Radiance Fields},
  booktitle = {Proc. IEEE/CVF Conf. Comput. Vis. Pattern Recognit. (CVPR)},
  year      = {2022}
}

@article{ZoeDepth2023,
  author  = {Bhat, Shariq Farooq and Birkl, Rico and Wofk, Diana and Wonka, Peter and M{\"u}ller, Matthias},
  title   = {ZoeDepth: Zero-shot Transfer by Combining Relative and Metric Depth},
  journal = {arXiv preprint arXiv:2302.12288},
  year    = {2023}
}

@inproceedings{TensoRF2022,
  author    = {Chen, Anpei and Xu, Zexiang and Ge, Fangcheng and others},
  title     = {TensoRF: Tensorial Radiance Fields},
  booktitle = {Proc. Eur. Conf. Comput. Vis. (ECCV)},
  year      = {2022}
}

@article{StableGS2024,
  author  = {Chen, Zigan and Li, Xi and Zhang, Yixing and Huang, Kun and Wang, Jian},
  title   = {StableGS: A Floater-Free Framework for 3D Gaussian Splatting},
  journal = {arXiv preprint arXiv:2403.18458},
  year    = {2024}
}

@inproceedings{BundleFusion2017,
  author    = {Dai, Angela and Nie{\ss}ner, Matthias and Zollh{\"o}fer, Michael and Izadi, Shahram and Theobalt, Christian},
  title     = {BundleFusion: Real-time Globally Consistent 3D Reconstruction using On-the-fly Surface Re-integration},
  booktitle = {ACM Trans. Graph. (SIGGRAPH)},
  year      = {2017}
}

@inproceedings{DepthSupervisedNeRF2022,
  author    = {Deng, Kang and Liu, Andrew and Springenberg, Jost Tobias and Brox, Thomas},
  title     = {Depth-supervised NeRF: Fewer Views and Faster Training for Free},
  booktitle = {Proc. IEEE/CVF Conf. Comput. Vis. Pattern Recognit. (CVPR)},
  year      = {2022}
}

@inproceedings{Plenoxels2022,
  author    = {Fridovich-Keil, Sara and Yu, Alex and Tancik, Matthew and Yang, Qianqian and Mildenhall, Ben and Kanazawa, Angjoo},
  title     = {Plenoxels: Radiance Fields Without Neural Networks},
  booktitle = {Proc. IEEE/CVF Conf. Comput. Vis. Pattern Recognit. (CVPR)},
  year      = {2022}
}

@article{Gauss2Mesh2024,
  author  = {Gu{\'e}don, Antoine and Lepetit, Vincent},
  title   = {Extracting Triangle Meshes from 3D Gaussians},
  journal = {arXiv preprint arXiv:2403.17868},
  year    = {2024}
}

@inproceedings{SuGaR2024,
  author    = {Gu{\'e}don, Antoine and Lepetit, Vincent},
  title     = {SuGaR: Surface-Aligned Gaussian Splatting for Efficient 3D Mesh Reconstruction and High-Quality Mesh Rendering},
  booktitle = {Proc. IEEE/CVF Conf. Comput. Vis. Pattern Recognit. (CVPR)},
  year      = {2024}
}

@article{HuberRobust1964,
  author  = {Huber, Peter J.},
  title   = {Robust Estimation of a Location Parameter},
  journal = {Ann. Math. Statist.},
  volume  = {35},
  number  = {1},
  pages   = {73--101},
  year    = {1964}
}

@article{ScreenedPoisson2013,
  author  = {Kazhdan, Michael and Hoppe, Hugues},
  title   = {Screened Poisson Surface Reconstruction},
  journal = {ACM Trans. Graph.},
  volume  = {32},
  number  = {3},
  pages   = {1--13},
  year    = {2013}
}

@inproceedings{UncertaintyWeighting2018,
  author    = {Kendall, Alex and Gal, Yarin and Cipolla, Roberto},
  title     = {Multi-task Learning Using Uncertainty to Weigh Losses for Scene Geometry and Semantics},
  booktitle = {Proc. IEEE/CVF Conf. Comput. Vis. Pattern Recognit. (CVPR)},
  year      = {2018}
}

@article{3DGS2023,
  author  = {Kerbl, Bernhard and Kopanas, Georgios and Leimk{\"u}hler, Thomas and Drettakis, George},
  title   = {3D Gaussian Splatting for Real-Time Radiance Field Rendering},
  journal = {ACM Trans. Graph.},
  volume  = {42},
  number  = {4},
  year    = {2023}
}

@article{TanksnTemples2017,
  author    = {Knapitsch, Arno and Park, Jaesik and Zhou, Qian-Yi and Koltun, Vladlen},
  title     = {Tanks and Temples: Benchmarking Large-Scale Scene Reconstruction},
  journal   = {ACM Trans. Graph.},
  volume    = {36},
  number    = {4},
  year      = {2017},
}

@inproceedings{AlexNet2012,
  author    = {Krizhevsky, Alex and Sutsever, Ilya and Hinton, Geoffrey E},
  title     = {ImageNet Classification with Deep Convolutional Neural Networks},
  booktitle = {Adv. Neural Inf. Process. Syst. (NeurIPS)},
  year      = {2012}
}

@article{MicroSplatting2024,
  author  = {Lee, Jee Won and Lim, Hoyeon and Yang, Sooyeun and Choi, Jongseong Brad},
  title   = {Micro-Splatting: Multistage Isotropy-Informed Covariance Regularization Optimization for High-Fidelity 3D Gaussian Splatting},
  journal = {arXiv preprint arXiv:2404.05740},
  year    = {2024}
}

@article{GS-Reg2024,
  author  = {Lee, Jo-seong and Choe, Geon-woo and Park, Min-cheol and Lee, Sang-eun and Kim, Min-hwan},
  title   = {GS-Reg: 3D Gaussian Splatting with Geometric Regularization},
  journal = {arXiv preprint arXiv:2404.14927},
  year    = {2024}
}

@inproceedings{NeRF2020,
  author    = {Mildenhall, Ben and Srinivasan, Pratul P. and Tancik, Matthew and Barron, Jonathan T. and Ramamoorthi, Ravi and Ng, Ren},
  title     = {NeRF: Representing Scenes as Neural Radiance Fields for View Synthesis},
  booktitle = {Proc. Eur. Conf. Comput. Vis. (ECCV)},
  year      = {2020}
}

@article{InstantNGP2022,
  author  = {M{\"u}ller, Thomas and Evans, Alex and Schied, Christoph and Keller, Alexander},
  title   = {Instant Neural Graphics Primitives with a Multiresolution Hash Encoding},
  journal = {ACM Trans. Graph.},
  volume  = {41},
  number  = {4},
  year    = {2022}
}

@inproceedings{KinectFusion2011,
  author    = {Newcombe, Richard A. and Izadi, Shahram and Hilliges, Otmar and Molyneaux, David and Kim, David and Davison, Andrew J. and Kohi, Pushmeet and Shotton, Jamie and Hodges, Steve and Fitzgibbon, Andrew},
  title     = {KinectFusion: Real-Time Dense Surface Reconstruction and Interaction},
  booktitle = {ACM SIGGRAPH Asia},
  year      = {2011}
}

@inproceedings{DPT2021,
  author    = {Ranftl, Ren{\'e} and Bochkovskiy, Alexey and Koltun, Vladlen},
  title     = {Vision Transformers for Dense Prediction},
  booktitle = {Proc. IEEE/CVF Int. Conf. Comput. Vis. (ICCV)},
  year      = {2021}
}

@article{MiDaS2022,
  author  = {Ranftl, Ren{\'e} and Lasinger, Katrin and Hafner, David and Schindler, Konrad and Koltun, Vladlen},
  title   = {Towards Robust Monocular Depth Estimation: Mixing Datasets for Zero-shot Cross-dataset Transfer},
  journal = {IEEE Trans. Pattern Anal. Mach. Intell.},
  volume  = {44},
  number  = {1},
  pages   = {382--397},
  year    = {2022}
}

@inproceedings{QSplat2000,
  author    = {Rusinkiewicz, Szymon and Levoy, Marc},
  title     = {QSplat: A Multiresolution Point Rendering System for Large Meshes},
  booktitle = {Proc. ACM SIGGRAPH},
  year      = {2000}
}

@inproceedings{SfMRevisited2016,
  author    = {Sch{\"o}nberger, Johannes Lutz and Frahm, Jan-Michael},
  title     = {Structure-from-Motion Revisited},
  booktitle = {Proc. IEEE/CVF Conf. Comput. Vis. Pattern Recognit. (CVPR)},
  year      = {2016}
}

@article{ScaffoldGS2024,
  author  = {Shao, Linning and Li, Le and Liao, Jin and Zhang, Wen and Zhang, Zhe and Liu, Hong and Wang, Guoping},
  title   = {Scaffold-GS: Structured 3D Gaussians for View-Adaptive Rendering},
  journal = {arXiv preprint arXiv:2403.16649},
  year    = {2024}
}

@inproceedings{DVGO2022,
  author    = {Sun, Cheng and Wang, Min and Jiang, Bin and Chen, Wei and Gao, Jia},
  title     = {Direct Voxel Grid Optimization: Super-fast Convergence for Radiance Fields Reconstruction},
  booktitle = {Proc. IEEE/CVF Conf. Comput. Vis. Pattern Recognit. (CVPR)},
  year      = {2022}
}

@inproceedings{Tancik_Nerfstudio_A_2023,
  author = {Tancik, Matthew and Reiser, Ethan and Mildenhall, Ben and others},
  title = {Nerfstudio: A Modular Framework for Neural Radiance Field Development},
  booktitle = {SIGGRAPH Asia 2023 Conference Papers},
  year = {2023},
}

@inproceedings{RefNeRF2022,
  author    = {Verbin, Daniel and Hedman, Peter and Mildenhall, Ben and Barron, Jonathan T. and Srinivasan, Pratul P.},
  title     = {Ref-NeRF: Structured View-Dependent Appearance for Neural Radiance Fields},
  booktitle = {Proc. IEEE/CVF Conf. Comput. Vis. Pattern Recognit. (CVPR)},
  year      = {2022}
}

@inproceedings{NeuS2021,
  author    = {Wang, Peng and Liu, Lingjie and Liu, Yaru and Theobalt, Christian and Komura, Taku and Wang, Wenping},
  title     = {NeuS: Learning Neural Implicit Surfaces by Volume Rendering for Multi-view Reconstruction},
  booktitle = {Adv. Neural Inf. Process. Syst. (NeurIPS)},
  year      = {2021}
}

@article{LowFreqFirst2023,
  author  = {Wang, Yipeng and Liu, Jiaming and Zhang, Kuan and Chen, Zhi-Hao and Yu, Peter and Xu, Lan},
  title   = {Low-Frequency First: Eliminating Floating Artifacts in 3D Gaussian Splatting},
  journal = {arXiv preprint arXiv:2308.02493},
  year    = {2023}
}

@inproceedings{FourDGS2024,
  author    = {Wu, Guanjun and Yi, Tian and Fang, Jiayi and Xie, Linyan and Zhang, Xinyu and Wei, Wei and Liu, Weijie and Tian, Qi and Wang, Xiaolong},
  title     = {4D Gaussian Splatting for Real-Time Dynamic Scene Rendering},
  booktitle = {Proc. IEEE/CVF Conf. Comput. Vis. Pattern Recognit. (CVPR)},
  year      = {2024}
}

@article{Survey3DGS2024,
  author  = {Wu, Tianle and Liu, Zhaoxi and Xu, Yifan and Wang, Yizhou},
  title   = {Recent Advances in 3D Gaussian Splatting},
  journal = {Computational Visual Media},
  year    = {2024}
}

@inproceedings{DepthAnything2024,
  author    = {Yang, Lihe and Kang, Bingyi and Huang, Zilong and Xu, Xiaogang and Feng, Jiashi and Zhao, Hengshuang},
  title     = {Depth Anything: Unleashing the Power of Large-Scale Unlabeled Data},
  booktitle = {Proc. IEEE/CVF Conf. Comput. Vis. Pattern Recognit. (CVPR)},
  year      = {2024}
}

@article{DepthAnythingV2_2024,
  author    = {Yang, Lihe and Kang, Bingyi and Huang, Zilong and Zhao, Zhen and Xu, Xiaogang and Feng, Jiashi and Zhao, Hengshuang},
  title     = {Depth Anything V2},
  journal = {arXiv preprint arXiv:2406.09414},
  year    = {2024}
}

@inproceedings{PlenOctrees2021,
  author    = {Yu, Alex and Ye, Ronghang and Tancik, Matthew and others},
  title     = {PlenOctrees for Real-time Rendering of Neural Radiance Fields},
  booktitle = {Proc. IEEE/CVF Int. Conf. Comput. Vis. (ICCV)},
  year      = {2021}
}

@inproceedings{MipSplatting2024,
  author    = {Yu, Zehao and Chen, Anpei and Huang, Bingliang and Sattler, Torsten and Geiger, Andreas},
  title     = {Mip-Splatting: Alias-free 3D Gaussian Splatting},
  booktitle = {Proc. IEEE/CVF Conf. Comput. Vis. Pattern Recognit. (CVPR)},
  year      = {2024}
}

@inproceedings{LPIPS2018,
  author    = {Zhang, Richard and Isola, Phillip and Efros, Alexei A. and Shechtman, Eli and Wang, Oliver},
  title     = {The Unreasonable Effectiveness of Deep Features as a Perceptual Metric},
  booktitle = {Proc. IEEE/CVF Conf. Comput. Vis. Pattern Recognit. (CVPR)},
  year      = {2018}
}

@inproceedings{LPGS2024,
  author    = {Zhang, Zhaoliang and Song, Tianchen and Lee, Yongjae and Yang, Li and Peng, Cheng and Chellappa, Rama and Fan, Deliang},
  title     = {LP-3DGS: Learning to Prune 3D Gaussian Splatting},
  booktitle = {Adv. Neural Inf. Process. Syst. (NeurIPS)},
  year      = {2024}
}

@inproceedings{PixelGS2024,
  author    = {Zhang, Hao and Wang, Yifan and Chen, Tianjian and Van Gool, Luc},
  title     = {Pixel-GS: Density Control with Pixel-aware Gradient for 3D Gaussian Splatting},
  booktitle = {Proc. Eur. Conf. Comput. Vis. (ECCV)},
  year      = {2024}
}

@inproceedings{NICE-SLAM2022,
  author    = {Zhu, Zihan and Peng, Songyou and Larsson, Viktor and Xu, Weiwei and Bao, Hujun and Cui, Zhaopeng and Pollefeys, Marc},
  title     = {{NICE-SLAM}: Neural Implicit Scalable Encoding for {SLAM}},
  booktitle = {Proc. IEEE/CVF Conf. Comput. Vis. Pattern Recognit. (CVPR)},
  year      = {2022}
}

@inproceedings{SurfaceSplatting2001,
  author    = {Zwicker, Matthias and Pfister, Hanspeter and van Baar, Jeroen and Gross, Markus},
  title     = {Surface Splatting},
  booktitle = {Proc. ACM SIGGRAPH},
  year      = {2001}
}

@ARTICLE{EWASplatting2002,
  author={Zwicker, M. and Pfister, H. and van Baar, J. and Gross, M.},
  journal={IEEE Transactions on Visualization and Computer Graphics}, 
  title={EWA splatting}, 
  year={2002},
  volume={8},
  number={3},
  pages={223-238},
  doi={10.1109/TVCG.2002.1021576}}

@article{ZipNeRF2023,
    title={Zip-NeRF: Anti-Aliased Grid-Based Neural Radiance Fields},
    author={Jonathan T. Barron and Ben Mildenhall and 
            Dor Verbin and Pratul P. Srinivasan and Peter Hedman},
    journal={ICCV},
    year={2023}
}

\newpage
\begin{IEEEbiography}[{\includegraphics[width=1in,height=1.25in,clip,keepaspectratio]{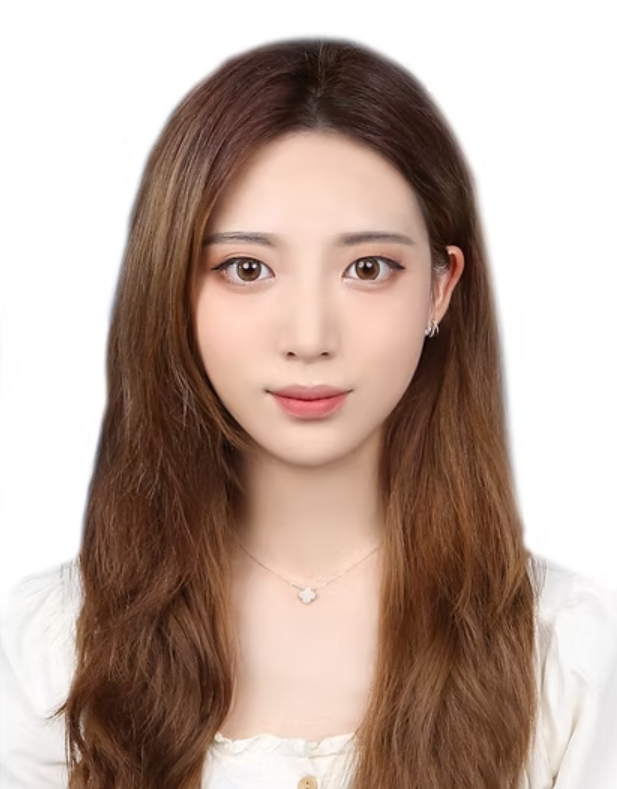}}]{Sooyeun Yang}
    received the BE degree in Mechanical Engineering from SUNY Korea, in 2024. Since then, She has been working toward the Master degree with the Department of Mechanical Engineering, SUNY Korea. Her research focuses on Digital Twin, Vision Sensing and Control, Radiance Field Rendering, and 3D Reconstruction.
\end{IEEEbiography}

\begin{IEEEbiography}[{\includegraphics[width=1in,height=1.25in,clip,keepaspectratio]{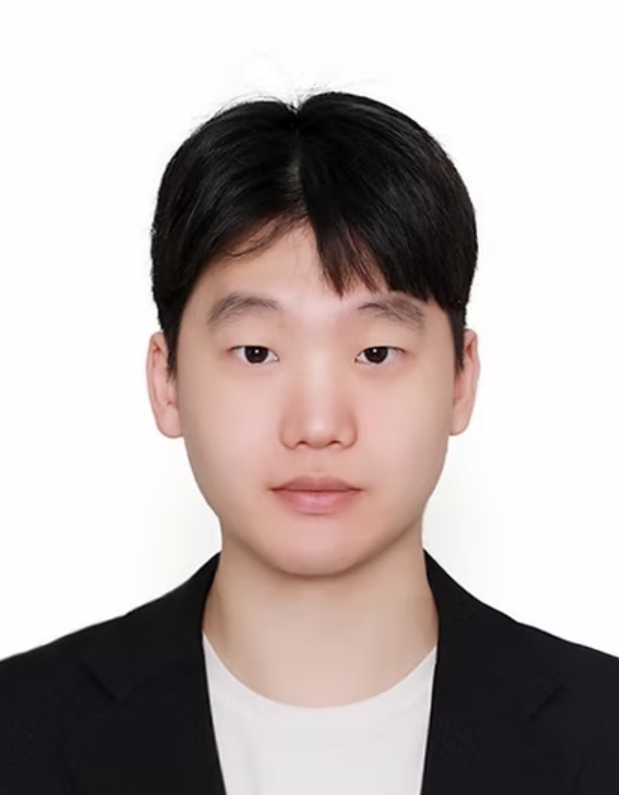}}]{Cheyul Im}
    received the BE degree in Mechanical Engineering from The State University of New York (SUNY) Korea, in 2025. He has been working toward the Master degree with the Department of Mechanical Engineering, SUNY Korea. His research interest is Physics-Informed Machine Learning, Computer Vision, and Digital Twin. 
\end{IEEEbiography}

\begin{IEEEbiography}[{\includegraphics[width=1in,height=1.25in,clip,keepaspectratio]{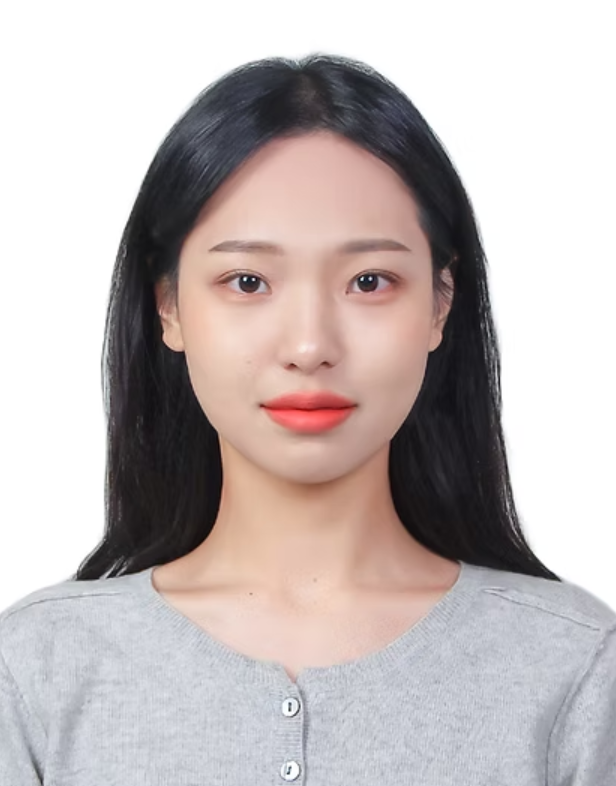}}]{Jee Won Lee}
    received her BE degree in Mechanical Engineering from The State University of New York (SUNY) Korea in 2021 and continued her her MS degree in Mechanical Engineering from SUNY Korea in 2025. She is currently working toward the Ph.D. degree with the Department of Mechanical Engineering, SUNY Korea. Her research interest is AI, Computer Vision, Embedded System, Degital Twin, and Deep Learning. 
\end{IEEEbiography}

\begin{IEEEbiography}[{\includegraphics[width=1in,height=1.25in,clip,keepaspectratio]{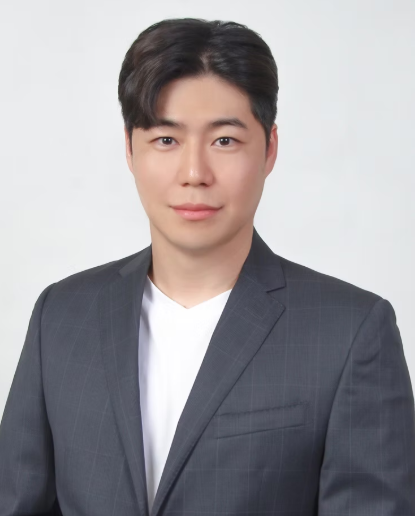}}]{Jongseong Brad Choi}
    received the Ph.D. degree in Mechanical Engineering from Purdue University, West Lafayette, IN, USA, in 2020. He is currently an Assistant Professor with the Department of Mechanical Engineering at The State University of New York (SUNY) Korea and holds a joint appointment as a Research Assistant Professor at Stony Brook University, USA. Prior to this, he was a Postdoctoral Researcher at Argonne National Laboratory. He has authored and coauthored numerous journal and conference papers in leading venues across computer vision, visual analytics, digital twins, and structural health monitoring. He has served as an organizer, session chair, and committee member for multiple international conferences. His research interests include visual analytics, computer vision, digital twin technologies, and human-machine collaborative systems for predictive engineering and infrastructure assessment. 
\end{IEEEbiography}

\vfill

\end{document}